\documentclass[%
aip,
amsmath,amssymb,
reprint,%
]{revtex4-2}
\usepackage{multirow}
\usepackage{graphicx}
\usepackage{dcolumn}
\usepackage{calc}
\usepackage{bm}
\usepackage[utf8]{inputenc}
\usepackage[T1]{fontenc}
\usepackage{mathptmx}
\usepackage{etoolbox}
\usepackage{xcolor}
\usepackage[table]{xcolor}
\usepackage{soul}
\usepackage[colorlinks=true,linkcolor=blue,citecolor=blue]{hyperref}%
\usepackage{amsfonts,amsmath,amssymb,amsthm,relsize}
\usepackage{hyperref}
\def\beq{\begin{equation}}
\def\eeq{\end{equation}}
\def\bea{\begin{eqnarray}}
\def\eea{\end{eqnarray}}

\makeatletter
\def\@email#1#2{%
\endgroup
\patchcmd{\titleblock@produce}
{\frontmatter@RRAPformat}
{\frontmatter@RRAPformat{\produce@RRAP{*#1\href{mailto:#2}{#2}}}\frontmatter@RRAPformat}
{}{}
}%
\makeatother
\newcommand{\kh}

\begin{document}

\title{Forecasting precipitation in the Arctic using probabilistic machine learning informed by causal climate drivers}

\newcommand{\daggerfootnote}{\textsuperscript{\textdagger}}

\author{Madhurima Panja}\thanks{These authors contributed equally to this work.}
\affiliation{Department of Science and Engineering, Sorbonne University, Abu Dhabi, United Arab Emirates}
\author{Dhiman Das}\thanks{These authors contributed equally to this work.}
\affiliation{Physics and Applied Mathematics Unit, Indian Statistical Institute, Kolkata 700108, India}

\author{Tanujit Chakraborty}
\affiliation{Department of Science and Engineering, Sorbonne University, Abu Dhabi, United Arab Emirates}
\affiliation{Sorbonne Center for Artificial Intelligence, Sorbonne University, Paris, 75005, France}

\author{Arnob Ray}
\affiliation{Indian Institute of Technology Gandhinagar, Gandhinagar 382355, India}

\author{R. Athulya}
\affiliation{National Centre for Polar and Ocean Research, Ministry of Earth Sciences, Vasco da Gama, 403804, India}

\author{Chittaranjan Hens}
\affiliation{Center for Computational Natural Science and Bioinformatics, International Institute of Information Technology, Gachibowli, Hyderabad-500032, India}

\author{Syamal K. Dana}
\affiliation{Department of Dynamics, Lodz University of Technology, Stefanowskiego 1/15, 90–537 Lodz, Poland}
\affiliation{Centre for Mathematical Biology and Ecology, Department of Mathematics, Jadavpur University, Kolkata 700032, India}

\author{Nuncio Murukesh}
\affiliation{National Centre for Polar and Ocean Research, Ministry of Earth Sciences, Vasco da Gama, 403804, India}

\author{Dibakar Ghosh}
\email{diba.ghosh@gmail.com}
\affiliation{Physics and Applied Mathematics Unit, Indian Statistical Institute, Kolkata 700108, India}

\begin{abstract}

Understanding and forecasting precipitation events in the Arctic maritime environments, such as Bear Island and Ny-{\AA}lesund, is crucial for assessing climate risk and developing early warning systems in vulnerable marine regions. This study proposes a probabilistic machine learning framework for modeling and predicting the dynamics and severity of precipitation. We begin by analyzing the scale-dependent relationships between precipitation and key atmospheric drivers (e.g., temperature, relative humidity, cloud cover, and air pressure) using wavelet coherence, which captures localized dependencies across time and frequency domains. To assess joint causal influences, we employ Synergistic-Unique-Redundant Decomposition, which quantifies the impact of interaction effects among each variable on future precipitation dynamics. These insights inform the development of data-driven forecasting models that incorporate both historical precipitation and causal climate drivers. To account for uncertainty, we employ the conformal prediction method, which enables the generation of calibrated non-parametric prediction intervals. Our results underscore the importance of utilizing a comprehensive framework that combines causal analysis with probabilistic forecasting to enhance the reliability and interpretability of precipitation predictions in Arctic marine environments.

\end{abstract}
\maketitle

\begin{quotation}

The rapid climate change in the Arctic region is reshaping precipitation patterns, with significant implications for marine ecosystems, navigation safety, and community resilience.  Within the vulnerable Svalbard archipelago, the coastal sites of Bear Island and  Ny-{\AA}lesund are particularly sensitive to changes in meteorological and oceanic conditions, making precise precipitation forecasting both a scientific and societal priority. However, the complex and nonlinear interaction between precipitation and its atmospheric drivers presents significant challenges to conventional prediction systems. In this paper, we introduce a probabilistic machine learning framework for modeling and forecasting precipitation dynamics in Arctic marine environments. By integrating scale-resolved dependency analysis, joint causal influence assessment, boosting-based forecaster, and conformal prediction intervals, our approach enhances both the accuracy and interpretability of forecasts, thereby supporting informed decision-making in one of the most climate-sensitive regions.
\end{quotation}

\maketitle

\section{\label{sec:intro} Introduction}
Precipitation patterns play a crucial role in shaping regional livelihoods, agricultural productivity, and water availability. However, excessive rainfall can lead to devastating consequences such as flooding, crop destruction, and the spread of waterborne diseases \cite{kangalawe2017climate, huq2004mainstreaming}. With global climate change, the frequency and intensity of extreme precipitation events have increased, posing growing challenges worldwide. These changes are particularly prominent in the Arctic, where extreme precipitation influences several key components of the regional climate system, including river discharge into the Arctic Ocean and enhanced surface ice melt \cite{vihma2016atmospheric, walsh2020extreme}. Despite its significance, extreme precipitation events in the Arctic marine environment remain understudied, largely due to the scarcity of high-resolution observational data. This gap is critical, as it represents one of the least understood aspects of the Arctic hydrological cycle with far-reaching global implications \cite{albeverio2006extreme, portner2022climate}. Research on extreme events has progressed \cite{ghil2011extreme, farazmand2019extreme, mishra2020routes, ray2020understanding, chowdhury2021extreme, chowdhury2022extreme, das2024complexity} in the broader perspective of dynamical systems,  where their emergence and the complex nonlinear processes involved in their origin have been explored, underscoring the universality of such phenomena and providing a conceptual background for interpreting precipitation extremes. Within this broader framework, analyzing Arctic precipitation trends offers valuable insights into global climate change, as shifts in snowfall and rainfall trends contribute to glacier and ice cap melt, ocean circulation changes, disrupted weather systems, and sea level rise, posing severe risks for low-lying islands and coastal regions \cite{box2019key, landrum2020extremes}.  To advance the understanding of these processes, this study focuses on the distinct climatic conditions of Bear Island (Bj{\o}rn{\o}ya) and Ny-{\AA}lesund regions that capture the critical aspects of Arctic precipitation dynamics.

Bj{\o}rn{\o}ya, located in the western Barents Sea between the high and low Arctic, occupies a strategic position for understanding Arctic marine climate dynamics. Situated just north of the polar front, where the North Atlantic Current divides into the West Spitsbergen Current and the Barents Sea Branch, the island lies upstream of other Svalbard sites and at the confluence of contrasting oceanographic influences \cite{loeng1991features, loeng1997water, carroll2011pan}. Its location within the Arctic amplification zone further enhances its role as a natural observatory for Arctic–North Atlantic linkages, particularly as the Barents–Kara region undergoes a rapid sea ice decline that strongly modulates local moisture transport and precipitation regimes. Despite this strategic importance, Bear Island has often exhibited relatively modest deviations in surface air temperature from long-term averages, suggesting a more moderate response to circulation changes during both warm and cold periods \cite{lupikasza2021importance}.

In contrast, Ny-{\AA}lesund on Spitsbergen represents a high Arctic counterpart, where the amplification signal is considerably stronger. Long-term observations from meteorological stations and the Zeppelin Observatory reveal pronounced winter and spring warming, alongside rising atmospheric moisture and circulation shifts \cite{maturilli2017arctic, platt2021atmospheric}. These differences make Ny-{\AA}lesund a critical site for capturing the direct impact of Arctic amplification on local hydrological and cryospheric processes. Ecological and glaciological studies reinforce these key differences: Dendrochronological records highlight the vulnerability of Bear Island’s tundra ecosystems to climatic variability and extremes \cite{owczarek2020post}, while in Ny-{\AA}lesund accelerated glacier retreat \cite{rs13193845}, permafrost thaw \cite{Putkonen12011998}, and associated feedbacks have been documented. These sites together illustrate how transitional (Bear Island) and high Arctic (Ny-{\AA}lesund) environments respond to climate variability.

Despite their importance, comprehensive studies of precipitation dynamics in these regions remain sparse. For Bear Island, investigations have primarily focused on precipitation patterns \cite{das2025pattern}, thunderstorm activity \cite{czernecki2015atmospheric}, and polar lows \cite{nordeng1992most}, while for Ny-{\AA}lesund, the role of atmospheric moisture transport and cloud–radiation interactions in shaping precipitation is recognized \cite{maturilli2017arctic, wendisch2019arctic, Pernov2022, platt2021atmospheric} but remains poorly constrained in long-term models. Addressing these gaps is essential, as precipitation variability at both sites is driven by complex interactions among atmospheric drivers that operate across multiple temporal scales, with Ny-{\AA}lesund typically showing more amplified relationships due to stronger coupling between sea ice decline, cloud processes, and moisture fluxes \cite{asutosh2021, ATHULYA2023106989}.

To capture these intricate dynamics, this study adopts a multifaceted approach that integrates causal analysis with deep learning architectures. In this integrated framework, a localized, scale-dependent perspective is first achieved through wavelet coherence analysis, which examines the time-frequency relationships between precipitation and key climatic variables such as temperature, relative humidity, cloud cover, and air pressure \cite{torrence1998practical, grinsted2004application}. This approach identifies both transient and persistent co-movements, as well as phase relationships that may indicate lagged influences. This is particularly valuable in climate-sensitive regions like the Arctic, where atmospheric interactions unfold over multiple temporal scales \cite{stroeve2012arctic}. Despite its effectiveness in revealing pairwise dependencies across time and frequency domains, wavelet coherence is limited in its ability to capture the combined influence of multiple variables or disentangle their shared and independent contributions to precipitation variability. To address this, we employ the Synergistic-Unique-Redundant Decomposition (SURD) framework. The information theory-based SURD approach partitions the total mutual information between precipitation and its predictors into unique, redundant, synergistic, and leak components \cite{martinez2024decomposing}. This allows us to isolate the independent contribution of each variable, quantify overlapping information, and capture joint interactions that only emerge in combination. Additionally, SURD accounts for self-causality, enabling us to evaluate the influence of lagged precipitation observations on its future variability. Together, the wavelet-based approach and the SURD analysis provide a comprehensive understanding of the dynamic processes governing precipitation in the Arctic marine environment.

While causality analysis provides critical insights into the atmospheric mechanisms driving precipitation dynamics, translating these insights into accurate and operationally useful forecasts remains a key challenge, particularly in the Arctic. Reliable forecasts play a crucial role in informing early warning systems and emergency preparedness strategies. In recent years, researchers have explored a range of time series forecasting techniques to improve the accuracy of precipitation predictions \cite{benziane2024survey}. Traditional statistical approaches, such as the Autoregressive Integrated Moving Average (ARIMA) model, have been widely applied for forecasting regional precipitation in Chicago \cite{lai2020use} and Sylhet, Bangladesh \cite{bari2015forecasting}. However, the limitations of these models in capturing the nonlinear, multiscale, and complex nature of climatic processes have resulted in a shift toward more flexible, data-driven approaches. Tree-based ensemble methods like Random Forest have shown promising results in generating precipitation forecasts in the United States \cite{herman2018money}. Similarly, deep neural networks have gained attention in forecasting precipitation dynamics owing to their ability to approximate highly nonlinear functions \cite{hall1999precipitation}. For instance, Shi et al. \cite{shi2015convolutional} introduced a convolutional recurrent neural network for nowcasting precipitation in Hong Kong, effectively capturing long-range temporal dependencies. Das et al.\ \cite{das2024hybrid} demonstrated that a physics–AI hybrid model outperforms numerical weather prediction for nowcasting extreme precipitation. More recently, attention-based architectures such as Transformers have demonstrated strong performance in modeling precipitation patterns in Xinjiang, benefiting from their capacity for parallel processing and modeling of long-sequence dependencies \cite{xu2024pfformer}. 

Despite advancements in precipitation forecasting, the majority of studies have focused on temperate and tropical regions, with relatively few addressing the unique challenges of modeling precipitation dynamics in Arctic environments characterized by sparse data, rapidly changing conditions, and strong atmospheric variability. Moreover, most existing models rely solely on historical precipitation data and often ignore the influence of key atmospheric drivers such as temperature, humidity, cloud cover, and air pressure. The inability of the forecasting models to capture the complex interactions among these variables may lead to inaccurate predictions. Additionally, the majority of existing work emphasizes point forecasts, providing limited insight into the uncertainty associated with predicted values. In climate-sensitive regions like the Arctic, the lack of probabilistic forecasting diminishes the practical utility of precipitation predictions and hinders the development of effective early warning systems. To address these limitations, we integrate historical precipitation and climatic variables with data-driven models and incorporate probabilistic forecasting through conformal prediction. This integrated approach enhances both the accuracy and uncertainty quantification of precipitation forecasts for Bear Island and Ny-{\AA}lesund, while enabling a direct comparison of transitional versus high Arctic responses to common large-scale drivers. By incorporating the causal climatic drivers such as temperature, relative humidity, cloud cover, and air pressure, along with the lagged precipitation dynamics, our framework aims to model the intricate dependencies that influence precipitation variability. We evaluate a suite of data-driven forecasting models under two settings: one that uses only past precipitation observations and another that includes the exogenous atmospheric variables. This dual strategy allows us to assess the usefulness of incorporating auxiliary information and to identify model architectures best suited for operational forecasting in the Arctic environment. Our empirical results strongly validate the importance of integrating auxiliary atmospheric information. Across all evaluated data-driven forecasting models, ranging from linear baselines to complex neural network architectures, the inclusion of exogenous variables consistently enhances forecasting accuracy. Notably, due to the inherent data scarcity in Arctic environments, tree-based models such as XGBoost with causal climate drivers as exogenous inputs outperform advanced deep neural architectures in terms of both accuracy and stability. This suggests that robust, interpretable ensemble methods may offer a more practical and effective solution for short-term precipitation forecasting in Bear Island and Ny-{\AA}lesund. Furthermore, the integration of conformal prediction with the XGBoost framework enables the generation of reliable prediction intervals, enhancing its practical utility for uncertainty-aware decision-making in high-risk Arctic environments.

The remainder of this paper is organized as follows. Section~\ref{sec:data} outlines the geographical context of the Bear Island and Ny-{\AA}lesund stations and provides a detailed description of the climatic datasets used in this study. Section~\ref{sec:Causality} presents the causality analysis, examining both individual-level and joint-level interactions between precipitation and atmospheric drivers. Section~\ref{forecasting} discusses the statistical characteristics of the precipitation data, evaluates the forecasting performance of various data-driven models under both univariate and multivariate settings, and validates the statistical significance of the performance improvements. Finally, Section~\ref{conclusion} summarizes the key findings and outlines potential directions for future research.
 
\section{\label{sec:data} Study area and description of data}

\begin{figure}
        \centering
	\includegraphics[width=\columnwidth]{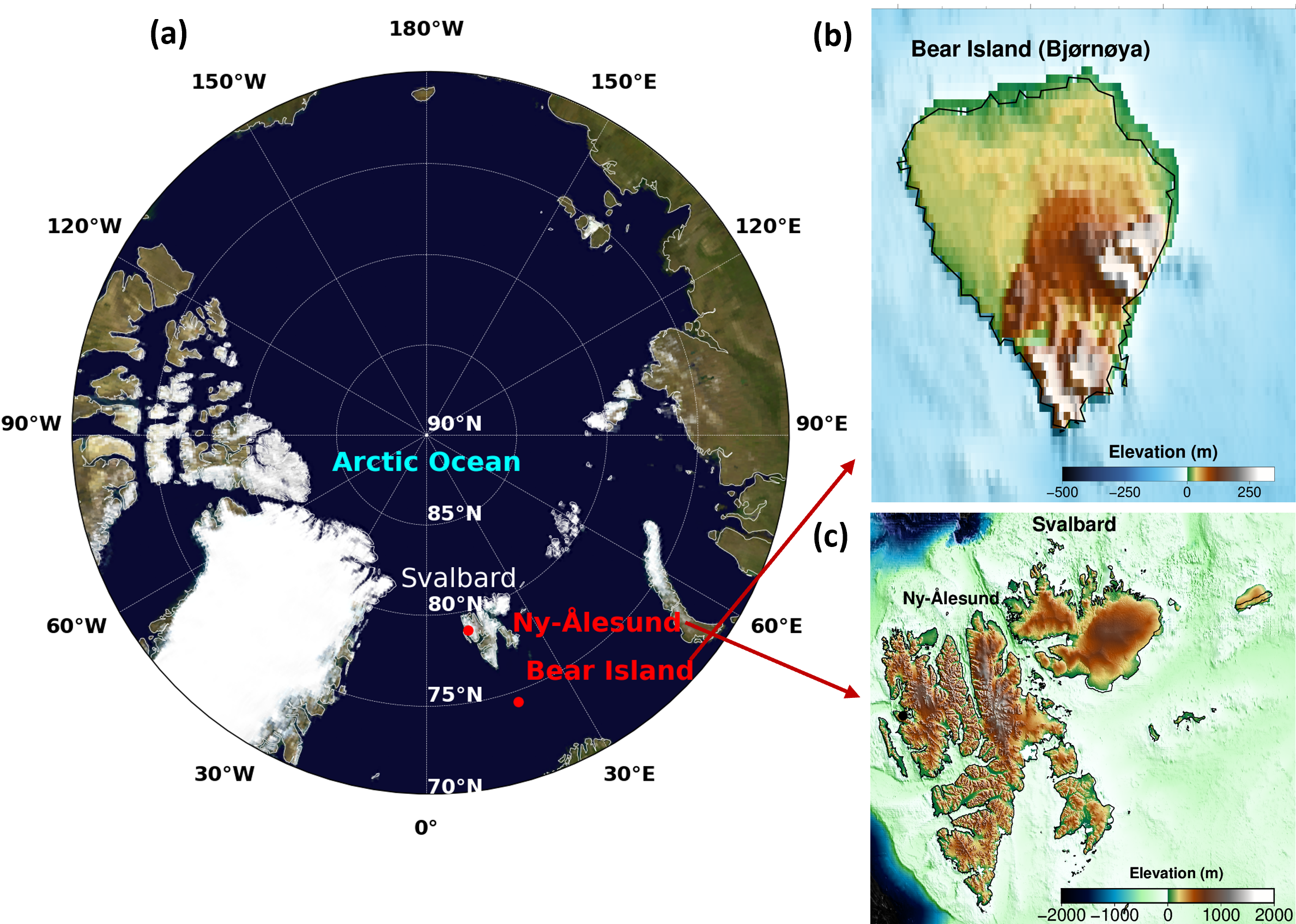}
	\caption{(a) Geographic locations of Bear Island (Bjørnøya; 74.4522° N, 19.1152° E) and  Ny-{\AA}lesund (78.923° N, 11.923° E), shown as red dots on an Arctic basemap generated with the NASA Blue Marble Earth image using Basemap v1.2.1. (b) High-resolution topographic map of Bear Island and (c)  Ny-{\AA}lesund, produced with PyGMT v0.16.0 and 15-arcsecond Earth relief data.}
	\label{fig0}
\end{figure}

We have collected daily observations of precipitation, mean air temperature, mean relative humidity, mean cloud cover, and average air pressure data from the meteorological station of Bj{\o}rn{\o}ya (station no. SN99710)  and Ny-{\AA}lesund (station no. SN99910) operated by the Norwegian Centre for Climate Services in Svalbard. The data span from January 1, 1991, to December 31, 2021, resulting in a dataset of 11,323 daily observations, free from missing entries.  Bj{\o}rn{\o}ya and Ny-{\AA}lesund are part of the  Arctic Svalbard archipelago. Bear Island is located in the western Barents Sea between mainland Norway and the North Pole, while Ny-{\AA}lesund is situated on the west coast of Spitsbergen, as shown in Figure \ref{fig0}. As the southernmost island in the archipelago, Bear Island spans approximately 20 km in length and up to 15 km in width. Its unique geography features flat and slightly hilly terrain in the northern and western regions, adorned with numerous lakes, while mountainous landscapes characterize the remainder of the island \cite{worslex2001geological}. The island experiences a maritime-polar climate heavily influenced by the North Atlantic current \cite{owczarek2020post}. Ny-{\AA}lesund, one of the northernmost settlements in the world at approximately 79°N latitude, lies along the southern shore of Kongsfjorden, surrounded by steep mountains and tidewater glaciers that drain into the fjord system \cite{Svendsen06012002}. Its high Arctic setting is characterized by a polar climate with long, cold winters and short, cool summers, strongly influenced by both Atlantic and Arctic air masses \cite{maturilli2017arctic}. The combination of fjord, glacier, and mountainous terrain makes Ny-{\AA}lesund a key observatory for atmosphere–cryosphere–ocean interactions in the Arctic. As precipitation dynamics in these regions often act as early indicators of broader shifts in atmospheric and oceanic systems, accurate forecasts can reveal circulation-driven variability at Bear Island and amplification-driven variability at Ny-{\AA}lesund, thereby providing critical insights for long-term decision making.

In this study, we consider the daily climatic time series and transform them into weekly aggregates to enable more robust modeling and forecasting of precipitation events. This temporal aggregation reduces short-term fluctuations in daily observations, which are often caused by localized weather variability, and instead emphasizes the persistent precipitation patterns that are climatologically significant. Compared to monthly averages, weekly aggregates are better at preserving extreme events while maintaining an optimal balance between capturing variability and avoiding excessive smoothing. Such weekly precipitation dynamics are therefore more relevant for long-term forecasting and the development of early warning systems in Arctic regions. Specifically, weekly total precipitation is obtained by summing daily observations, while weekly averages of temperature, humidity, cloud cover, and air pressure are computed using arithmetic means. To maintain consistency in the dataset, the final week of 2021, which contains only five days (December 27–31), is retained as a separate observation. 

\begin{figure*}
    \centering
    \includegraphics[width=0.9\linewidth]{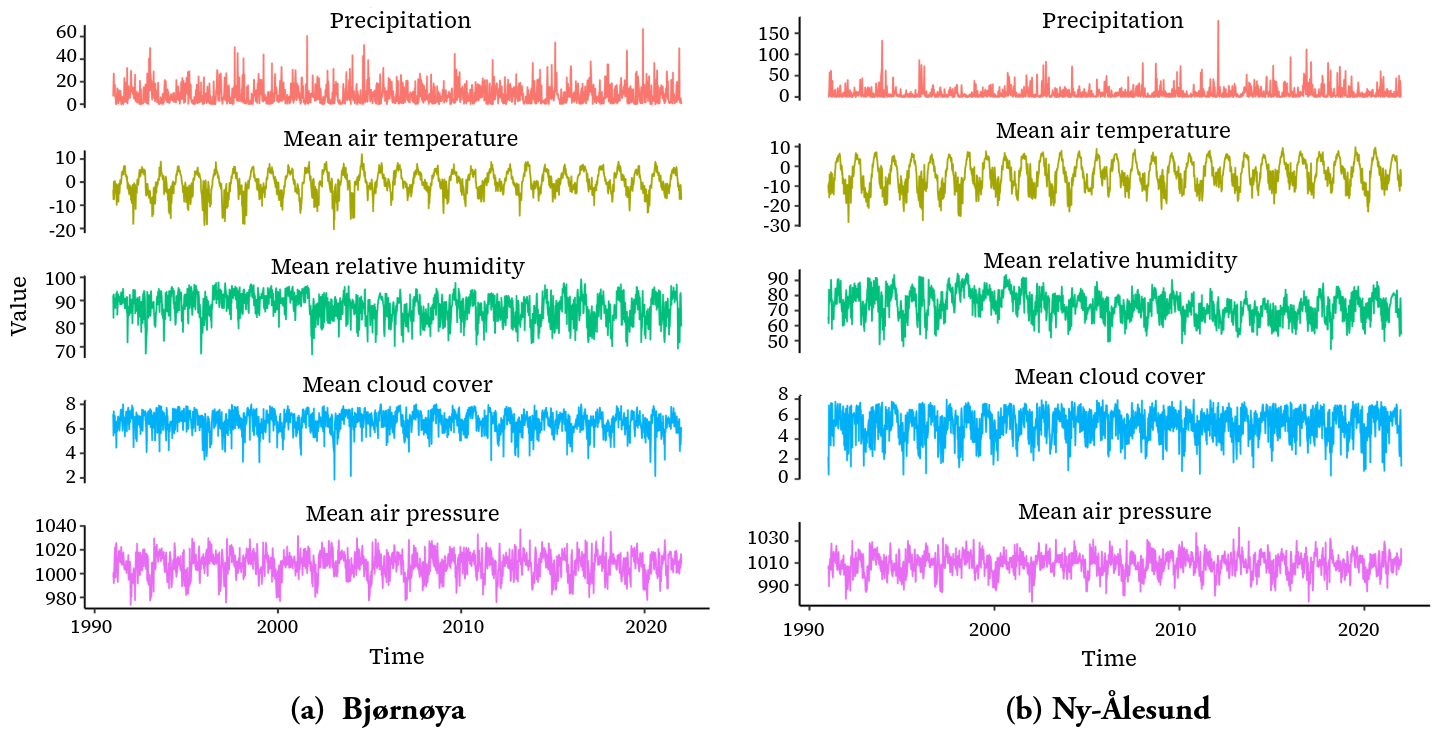}
    \caption{Temporal evolution of precipitation, mean air temperature, mean relative humidity, mean cloud cover, and average air pressure recorded in (a) Bj{\o}rn{\o}ya and (b) Ny-{\AA}lesund regions during 1991-2021. Weekly aggregated variables depict the seasonal trends and long-term variation of the series.}
    \label{Fig_Variable_Plots}
\end{figure*}

Figure \ref{Fig_Variable_Plots} illustrates the temporal behavior of the climatic variables observed at Bear Island and Ny-{\AA}lesund regions during 1991-2021 at a weekly frequency, capturing both seasonal trends and long-term variability in the dataset. As evident from the plot, precipitation events in the Arctic, although exhibiting irregular spikes throughout the year, are primarily associated with periods of increased relative humidity and elevated cloud cover, consistent with established meteorological principles \cite{richards1981relationship}. In contrast, higher air temperatures are generally associated with enhanced evaporation rates and increased vapor pressure deficits, which can restrict precipitation extremes \cite{wang2018dependence}. Similarly, increased air pressure often suppresses cloud formation and is typically linked to reduced precipitation \cite{yu2018bridge}. These observed relationships underscore the importance of incorporating atmospheric drivers into the modeling framework to improve the accuracy and interpretability of precipitation forecasts in Arctic environments.

\section{\label{sec:Causality}Causality analysis}
Understanding the relationships between precipitation and key climatic drivers is essential for characterizing regional climate dynamics and uncovering the mechanisms that govern precipitation events. In this section, we investigate these interactions across different time scales and levels of complexity. To capture individual-level dependencies, we employ wavelet coherence analysis, which reveals localized, scale-dependent relationships between precipitation and each of the climatic variables \cite{grinsted2004application}. Additionally, to examine joint-level interactions, we adopt the SURD framework, which quantifies how combinations of climatic factors jointly influence precipitation \cite{martinez2024decomposing}.

\subsection{Individual-level Interactions}
To investigate the interdependence between precipitation and the atmospheric variables across time and frequency domains, we employ wavelet coherence analysis \cite{grinsted2004application, panja2023ensemble}. This approach allows us to examine the scale-dependent co-movement between precipitation and four climatic drivers, such as temperature, relative humidity, cloud cover, and air pressure. By capturing localized coherence patterns over time, wavelet coherence provides valuable insights into their mutual influence, including potential lead-lag relationships that may influence precipitation events.

\begin{figure}
	\centerline{
		\includegraphics[width=\columnwidth]{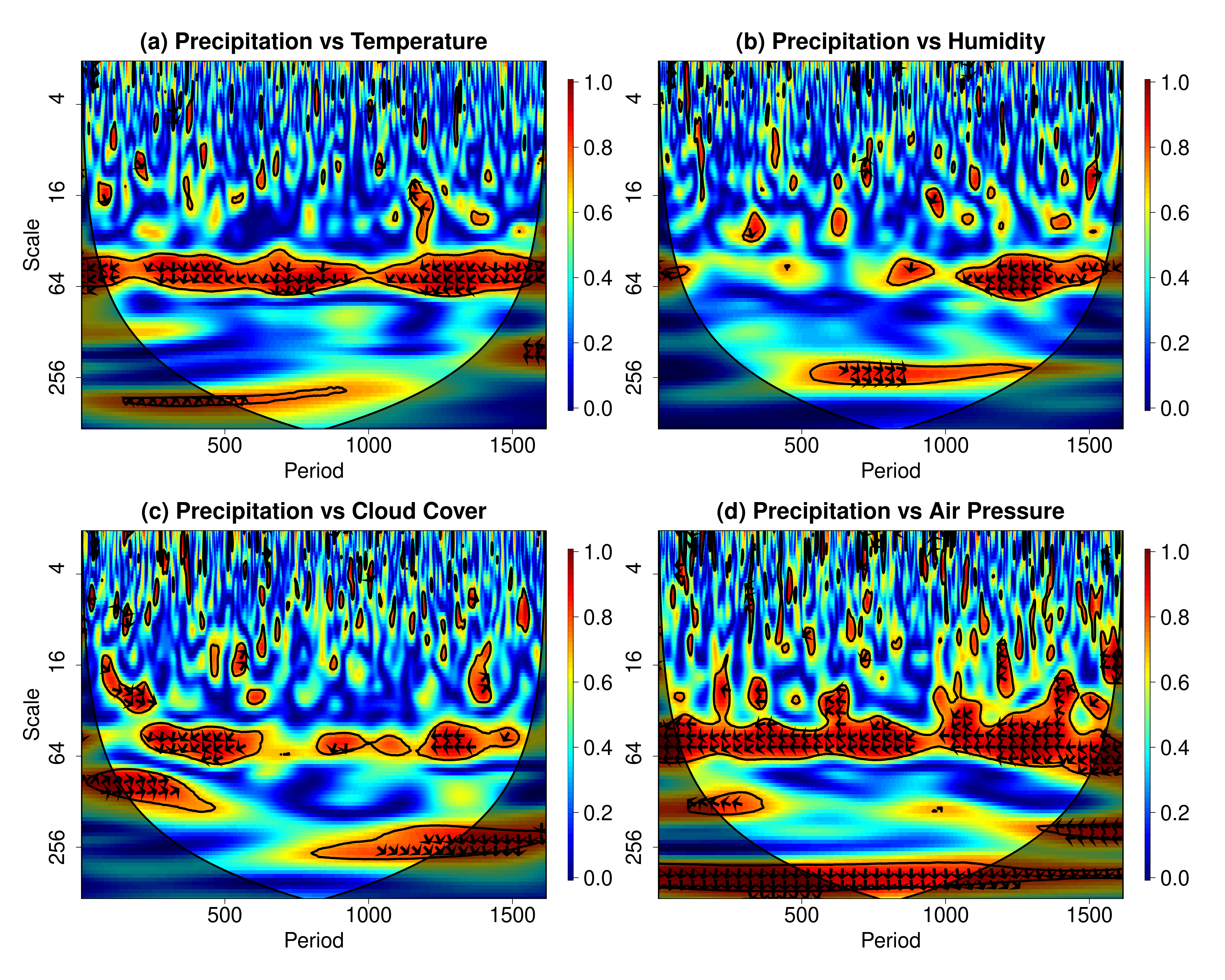}
	}
	\caption{Wavelet coherence analysis of precipitation and climatic variables at  Bj{\o}rn{\o}ya from 1991 to 2021. Figures (a) to (d) demonstrate the wavelet coherence between precipitation and each of the four climatic variables: temperature, relative humidity, cloud cover, and air pressure, respectively. Warmer hues (red/yellow) indicate strong coherence, while cooler colors (blue) denote weak coherence. Arrows represent phase relationships: rightward arrows indicate in-phase synchronization, leftward arrows show anti-phase interactions, and upward/downward arrows denote lead-lag dynamics.  Black contours highlight regions statistically significant at the 5\% level, tested against the null hypothesis that both time series are independent red-noise (AR(1)) processes. Significance was determined via Monte Carlo simulations, with contours marking areas where observed coherence exceeds the 95th percentile of surrogate distributions.} The gray area outside the cone of impact denotes regions affected by edge effects. The y-axis (log$_2$-transformed) represents temporal scales in weeks, showing both short-term 
(bottom) and long-term (top) patterns, while the x-axis shows the temporal progression of the 
31-year study period.
	\label{fig8}
\end{figure}

\begin{figure}
	\centerline{
		\includegraphics[width=\columnwidth]{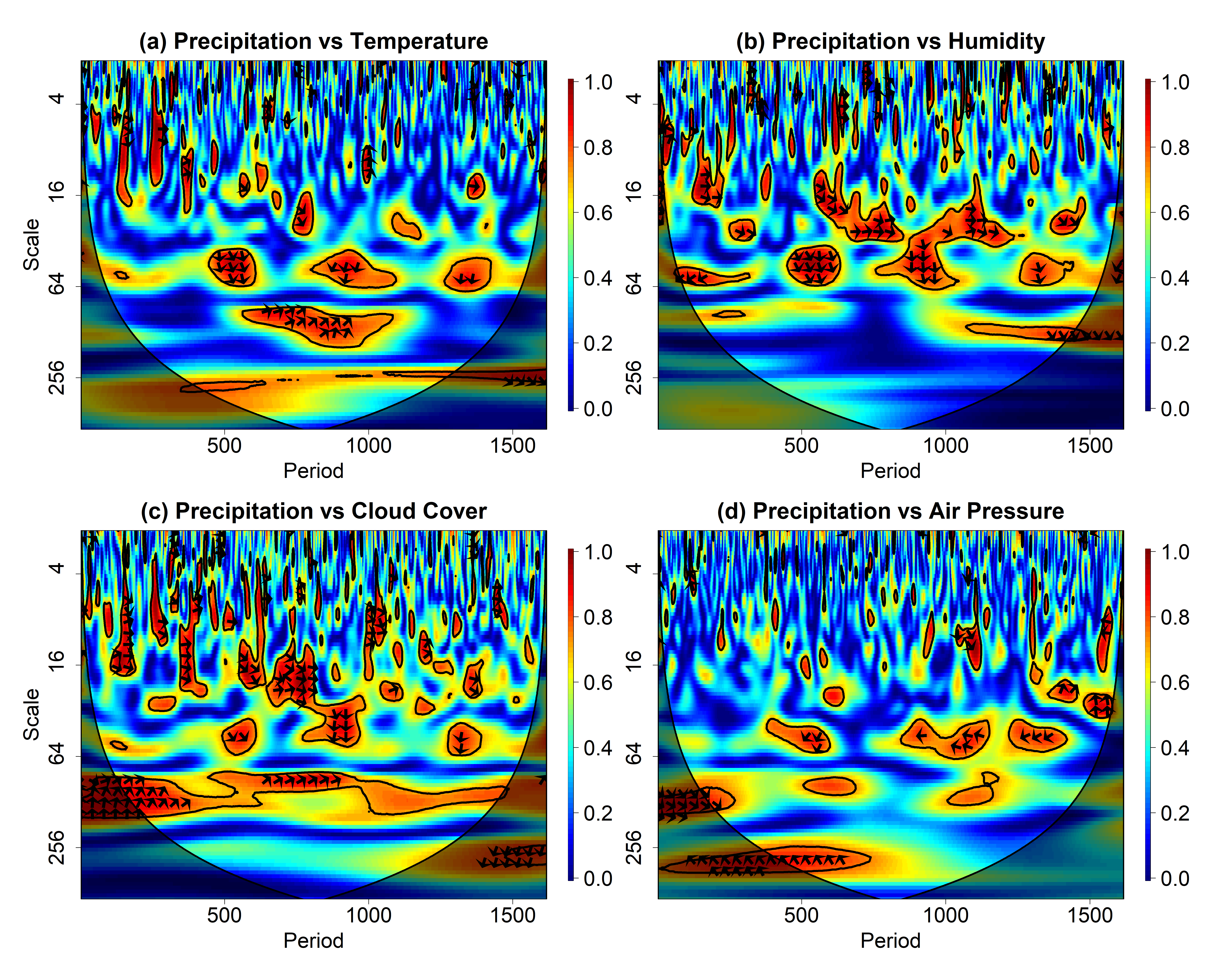}
	}
	\caption{Wavelet coherence analysis of precipitation and climatic variables at  Ny-{\AA}lesund from 1991 to 2021. Figures (a) to (d) demonstrate the wavelet coherence between precipitation and each of the four climatic variables: temperature, relative humidity, cloud cover, and air pressure, respectively. Warmer hues (red/yellow) indicate strong coherence, while cooler colors (blue) denote weak coherence. Arrows represent phase relationships: rightward arrows indicate in-phase synchronization, leftward arrows show anti-phase interactions, and upward/downward arrows denote lead-lag dynamics. Black contours highlight regions statistically significant at the 5\% level, tested against the null hypothesis that both time series are independent red-noise (AR(1)) processes. Significance was determined via Monte Carlo simulations, with contours marking areas where observed coherence exceeds the 95th percentile of surrogate distributions. The gray area outside the cone of impact denotes regions affected by edge effects. The y-axis (log$_2$-transformed) represents temporal scales in weeks, showing both short-term 
(bottom) and long-term (top) patterns, while the x-axis shows the temporal progression of the 
31-year study period.}
	\label{fig9}
\end{figure}

Figures \ref{fig8} and \ref{fig9} depict the wavelet coherence analysis of precipitation and each of the four climatic variables: temperature, relative humidity, cloud cover, and air pressure at  Bj{\o}rn{\o}ya and Ny-{\AA}lesund, respectively. The x-axis indicates the period (time scale), and the y-axis specifies the scale (frequency). The color scale, which ranges from blue to red, represents the strength of coherence, with red portions indicating strong coherence and blue sections suggesting low coherence. Black contours denote statistically significant coherence zones, whereas arrows within these regions suggest phase correlations. Rightward arrows indicate that the two signals are in phase, whereas leftward arrows indicate an out-of-phase relationship and upward or downward arrows denote phase lags or leads.

 At  Bj{\o}rn{\o}ya, precipitation-temperature (Figure \ref{fig8}a) and precipitation-cloud cover (Figure \ref{fig8}c) show significant coherence at seasonal scales, while precipitation-humidity (Figure \ref{fig8}b) exhibits strong, continuous coherence across all time scales. Precipitation-pressure (Figure \ref{fig8}d) shows weak, scattered coherence at low frequencies. In the case of Ny-{\AA}lesund, the wavelet coherence plot, as presented in Figure \ref{fig9}, depicts weaker and more fragmented patterns. Precipitation-temperature (Figure \ref{fig9}a) and precipitation-cloud cover (Figure \ref{fig9}c) depict the strongest coherence, but with less temporal continuity than on Bjørnøya. Precipitation-humidity (Figure \ref{fig9}b) coherence is barely observable, but precipitation-pressure (Figure \ref{fig9}d) coherence can be identified. Multi-scale atmospheric processes are responsible for these complicated patterns observed at both locations. The patchy coherence indicates threshold-dependent interactions, with linkages strengthening only under certain meteorological conditions. The continuous in-phase humidity connection at both locations indicates direct thermodynamic coupling, proving the critical role of atmospheric moisture in Arctic precipitation. The discrepancies between stations emphasize geographical influences: Bj{\o}rn{\o}ya's coastal position exhibits stronger ocean-influenced couplings, while  Ny-{\AA}lesund's fjord setting reveals more disordered patterns due to local topological impacts.

\subsection{Joint-level Interactions}
Wavelet coherence analysis reveals the time-frequency co-movements between precipitation and individual climatic variables; however, it is inherently limited to pairwise relationships and cannot capture the unique and interaction effects of multiple climatic drivers. To address this, we adopt the SURD framework, an information-theoretic approach that enables a structured decomposition of information flow from lagged precipitation and climatic variables to future precipitation values \cite{martinez2024decomposing}. This framework captures self-causation, nonlinear dependencies, stochastic effects, and collider influences, providing a more comprehensive understanding of multivariate interactions.

\begin{figure}
    \centering
    \includegraphics[width=\columnwidth]{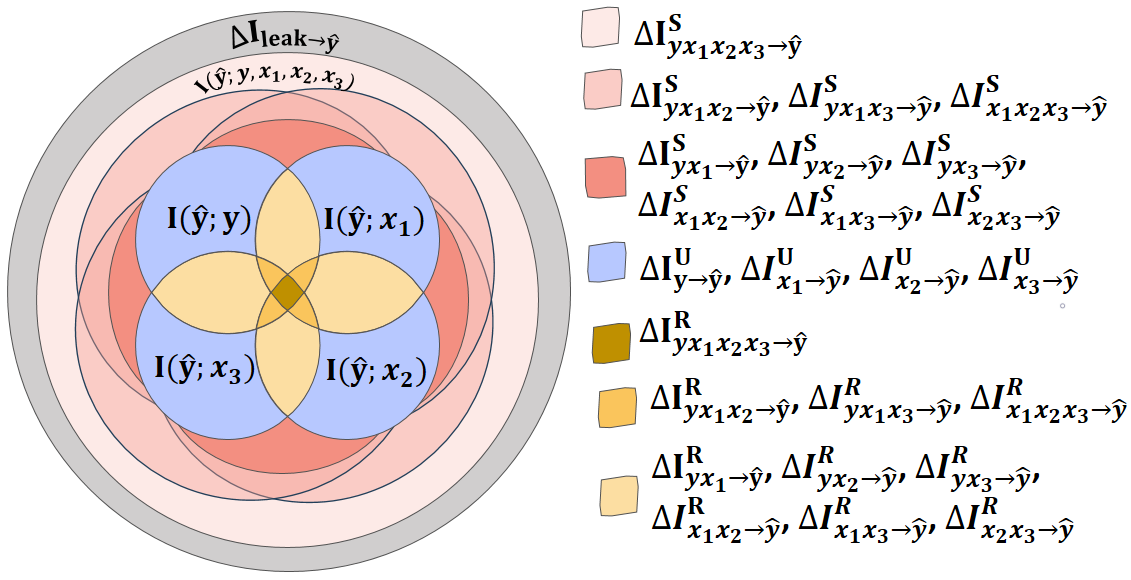}
    \caption{SURD analysis of causal information from lagged variables ${y, x_1, x_2, x_3}$ to the future target $\hat{y}$. The decomposition illustrates how the total $\mathrm{I}(\hat{y}; y, x_1, x_2, x_3 )$ and individual $\mathrm{I}(\hat{y}, \cdot)$ mutual information is distributed across synergistic (red), unique (blue), and redundant (yellow) components. The term $\Delta \mathrm{I}_{\text{leak} \rightarrow \hat{y}}$ (gray) captures unexplained or leaked causality. Specifically, $\Delta \mathrm{I}_{(\cdot) \rightarrow \hat{y}}^R$ quantifies redundant information shared among multiple predictors $(\cdot)$ about target $\hat{y}$, while $\Delta \mathrm{I}_{(\cdot) \rightarrow \hat{y}}^U$ and $\Delta \mathrm{I}_{(\cdot) \rightarrow \hat{y}}^S$ represent the unique and synergistic causal contributions, respectively, from the predictors to the target variable.}
    \label{Fig_SURD_Model}
\end{figure}

Given precipitation dynamics $y_t$, our goal is to assess the causal contributions of its past values and those of $N$ climatic variables $\mathbf{X}_t=\left\{x_{1, t}, x_{2, t}, \ldots, x_{N, t}\right\}$ to future outcomes $\hat{y}_{t+h}$ for a forecast horizon $h>0$. SURD quantifies causality as the increase in information $\left(\Delta \mathrm{I}\right)$ about $\hat{y}_{t+h}$ derived from observing individual or joint subsets of predictors in $\mathcal{X}_t=\left\{y_t, \mathbf{X}_t\right\}$. It decomposes the total Shannon entropy $\mathrm{H}\left(\hat{y}_{t+h}\right)$ into distinct causal components: redundant $\left(\Delta \mathrm{I}_{k \rightarrow \hat{y}}^R\right)$, unique $\left(\Delta \mathrm{I}_{k \rightarrow \hat{y}}^U\right)$, synergistic $\left(\Delta \mathrm{I}_{k \rightarrow \hat{y}}^S\right)$, and leak causality $\left(\Delta \mathrm{I}_{\text{leak}\rightarrow \hat{y}} \right)$ as
$$
\mathrm{H}(\hat{y}_{t+h}) = \sum_{k \in \bar{\mathcal{X}_t}} \Delta \mathrm{I}_{k \rightarrow \hat{y}}^R + \sum_{k \in \mathcal{X}_t} \Delta \mathrm{I}_{k \rightarrow \hat{y}}^U + \sum_{k \in \bar{\mathcal{X}_t}} \Delta \mathrm{I}_{k \rightarrow \hat{y}}^S + \Delta \mathrm{I}_{\text{leak} \rightarrow \hat{y}}, 
$$
where $\bar{\mathcal{X}_t}$ denotes all combinations of variables in $\mathcal{X}_t$. The non-negative causal components derived from SURD represent distinct associations where redundant causality $\left(\Delta \mathrm{I}_{k \rightarrow \hat{y}}^R\right)$ reflects overlapping information shared by multiple variables, unique causality $\left(\Delta \mathrm{I}_{k \rightarrow \hat{y}}^U\right)$ captures information solely attributed to individual variable, synergistic causality $\left(\Delta \mathrm{I}_{k \rightarrow \hat{y}}^S\right)$ arises from interactions that provide more information together than individually, and leak causality $\left(\Delta \mathrm{I}_{\text{leak} \rightarrow \hat{y}}\right)$ accounts for unobserved influences not captured by the observed set. Figure \ref{Fig_SURD_Model} presents this decomposition for $N=3$ as an illustrative example. In our analysis, we apply this framework using precipitation as the target variable and all four climatic drivers as inputs. Notably, SURD prevents duplication of causal contributions, ensuring that each component is precisely quantified and additive. The mutual information between the future target and the full predictor set $\mathrm{I}\left(\hat{y}_{t+h}; \mathcal{X}_t\right)$ is quantified as the sum of all observed causal components, while mutual information with individual predictors (say $x_{i, t} \in \mathcal{X}_t$), denoted as $\mathrm{I}\left(\hat{y}_{t+h}; x_{i, t}\right)$ is composed of their unique and redundant effects. To facilitate interpretability, these causal components are normalized by $\mathrm{I}\left(\hat{y}_{t+h}; \mathcal{X}_t\right)$, ensuring their sum equals one, whereas leak causality is normalized by $\mathrm{H}\left(\hat{y}_{t+h}\right)$, yielding values between 0 (fully explained) and 1 (entirely unexplained). 

\begin{figure*}
    \centering
    \includegraphics[width=\linewidth]{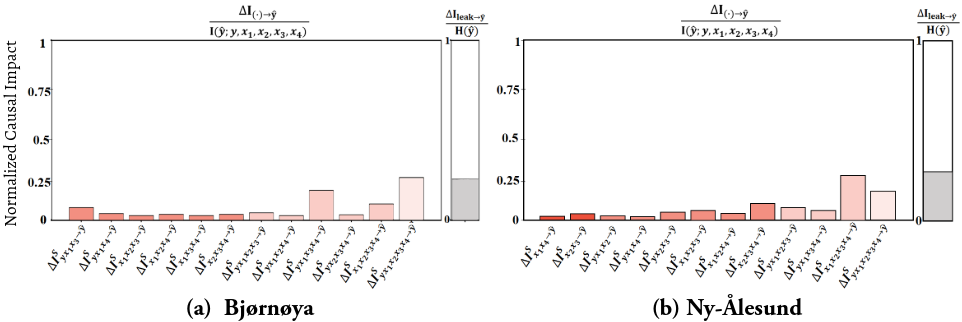}
    \caption{Schematic representation of the SURD analysis for (a) Bj{\o}rn{\o}ya and (b) Ny-{\AA}lesund regions with normalized causal values illustrating the causal influence on future precipitation values ($\hat{y}$) from its lagged observations ($y$) and climatic variables - temperature ($x_1$), relative humidity ($x_2$), cloud cover ($x_3$), and air pressure ($x_4$). The dominant synergistic causal interactions are highlighted with varying shades of red: lighter shades represent interactions involving a greater number of variables, while darker shades correspond to interactions among fewer variables. The length of each bar reflects the strength of the causal influence. The gray shaded bar quantifies the normalized leak causality arising from unobserved factors.}
    \label{Fig_SURD_Analysis}
\end{figure*}

Figure \ref{Fig_SURD_Analysis} visualizes the results of the SURD framework applied to precipitation dynamics in Bear Island and Ny-{\AA}lesund, using lagged precipitation and key climatic drivers, including temperature, relative humidity, cloud cover, and air pressure. Each bar represents the normalized causal contribution of a SURD component (unique, redundant, synergistic, or leak), with red shades denoting synergistic interactions and the gray bar quantifying leak causality from unobserved influences. For a better visualization, we present the twelve most influential components that significantly contribute to the future patterns of precipitation. The analysis reveals that for Bear Island, the synergistic components, capturing interaction effects, vividly explain 96.08\% of the total mutual information, with the joint interaction of all predictors contributing approximately 28\%. In contrast, the unique and redundant components play a minor role by respectively quantifying 1.28\% and 2.62\% of the overall mutual information. The leak causality component highlights that unobserved variables account for about 23\% of the overall causality, suggesting that the lagged values of precipitation, temperature, relative humidity, cloud cover, and air pressure together explain 77\% of the future precipitation variability in Bj{\o}rn{\o}ya. The SURD analysis on Ny-{\AA}lesund dataset exhibits a similar interaction-driven structure. The synergistic components explain 94.56\% of the total mutual information, with joint climatic drivers contributing about 25\%. Unique (2.46\%) and redundant (2.98\%) contributions remain limited, while 27\% of the information is captured by leak causality. This increased unexplained variability indicates that unobserved variables play a stronger role in shaping precipitation dynamics in Ny-Ålesund. Overall, the SURD reveals that the evolution of Arctic precipitation in both regions emerges from complex, nonlinear, multivariate interactions among climatic drivers. The increased synergistic components of the SURD framework underscore how the climatic variables interact to regulate condensation, cloud formation, and moisture transportation, particularly under the unique thermodynamic constraints of Arctic environments. Beyond aligning with established atmospheric theory, the SURD-based analysis introduces an information-theoretic perspective that quantifies the strength and nature of these coupled effects, providing a novel, data-driven interpretation of precipitation dynamics in the region. These findings serve as valuable prior knowledge for enhancing data-driven forecasting models. Integrating these atmospheric predictors can guide machine learning algorithms to generate reliable and accurate precipitation forecasts.

 \section{\label{forecasting} Forecasting Precipitation Dynamics}

Precipitation forecasting remains one of the most complex and essential tasks in climate science, driven by its crucial role in water resource management, agricultural planning, disaster mitigation, and ecosystem monitoring \cite{hall1999precipitation}. Unlike other climatic variables like temperature or pressure, precipitation is highly variable and often characterized by chaotic behavior, which is difficult to model using traditional approaches. Predicting the temporal evolution of precipitation dynamics requires not only accurate historical records but also a comprehensive understanding of the physical and climatic factors that drive its variability. In this study, we focus on generating probabilistic forecasts for the weekly evolution of precipitation dynamics in the Arctic region using machine learning approaches. The objective is to assess how the inclusion of key climatic drivers, such as temperature, relative humidity, cloud cover, and air pressure, impacts the forecasting capabilities of data-driven models. To evaluate both performance and robustness of our approach, we conduct empirical analyses on two distinct Arctic regions: Bear Island and Ny-{\AA}lesund. In both cases, forecasting models incorporate key climatic drivers, including temperature, relative humidity, cloud cover, and air pressure, identified in Section \ref{sec:Causality} as causal determinants of precipitation. These variables provide physically meaningful signals that enrich historical precipitation data and enhance the predictive capability of the models. In this analysis, we generate one-step-ahead forecasts for precipitation levels of Arctic regions, iteratively extending over a 53-week horizon. These weekly forecasts form a robust basis for understanding the variability of precipitation. The week-by-week predictions, generated by the forecasting models, capture the short-term evolution of precipitation patterns, thereby supporting operational decision-making, resource management, and preparedness for abrupt weather changes. Thus, by integrating causally impacting climatic drivers with data-driven probabilistic forecasting frameworks, this study offers a comprehensive approach for responsive short-term planning and developing early warning systems to mitigate precipitation changes in the Arctic. In the following sections, we discuss the global characteristics of the climatic variables, the forecasting techniques utilized in this study, the performance evaluation metrics, benchmark comparisons, and uncertainty quantification to assess the reliability of the predictions.

\subsection{Global Features}\label{Sec_Global_Features}

\begin{figure*}
    \centering
    \includegraphics[width=\linewidth]{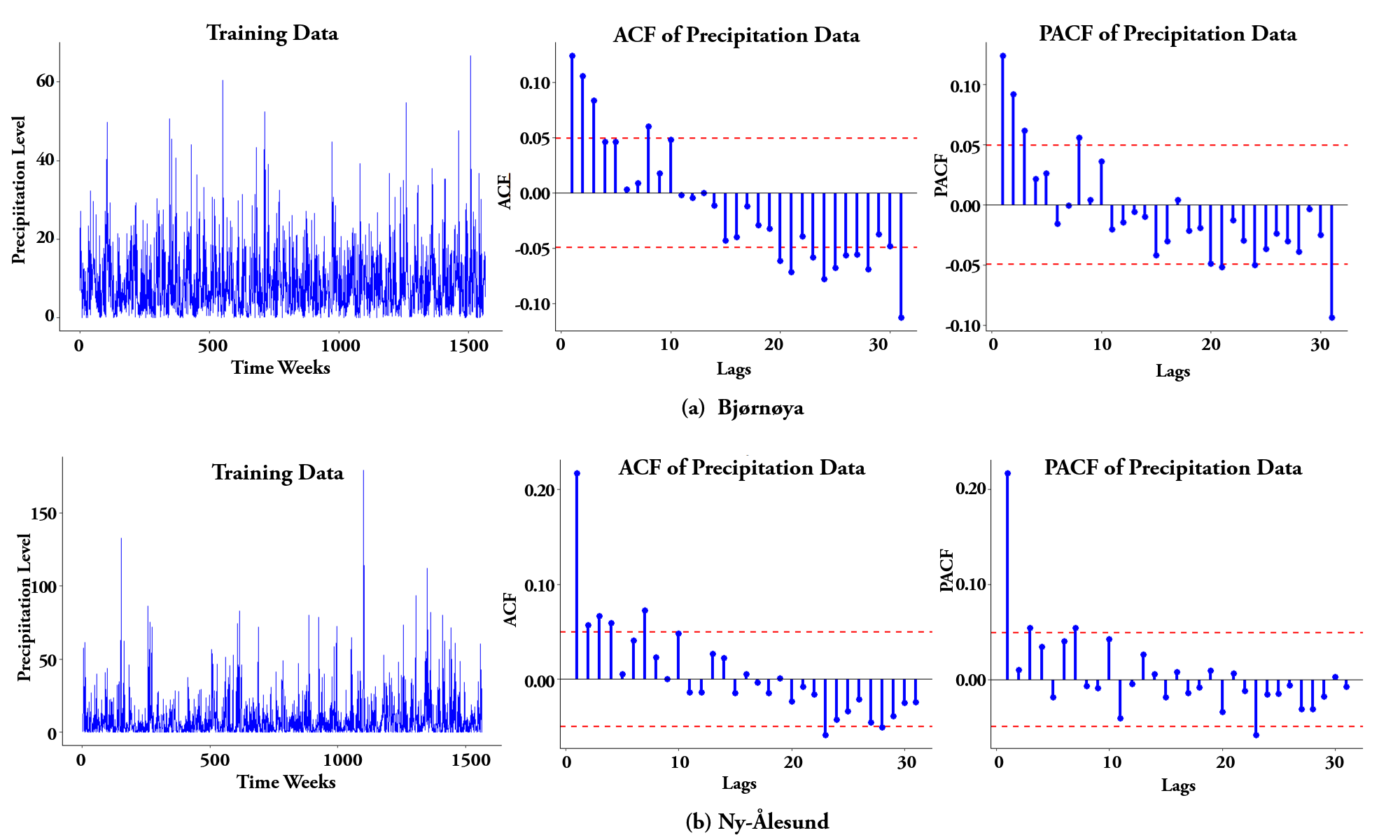}
    \caption{The plot visualizes the precipitation dynamics of (a) Bj{\o}rn{\o}ya and (b) Ny-{\AA}lesund regions used to train the data-driven forecasting models (left), along with the corresponding autocorrelation function (ACF) (middle) and partial autocorrelation function (PACF) (right).}
    \label{Fig_ACF_PACF_Plot}
\end{figure*}

In this section, we analyze the global characteristics of the weekly precipitation levels recorded in the Bj{\o}rn{\o}ya and Ny-{\AA}lesund regions over the study period. For empirical evaluation, the weekly datasets spanning from January 1, 1991, to December 31, 2021, are chronologically divided into two subsets: a training set comprising 1,565 weekly observations from January 1, 1991, to December 27, 2020, and a test set consisting of 53 weekly observations from January 3, 2021, to December 31, 2021. The training data for Bj{\o}rn{\o}ya records an average weekly precipitation level of 8.65 mm, with values ranging from a minimum of 0 mm to a maximum of 66.70 mm. The coefficient of variation, which measures relative dispersion around the mean, is 95.94\%, underscoring the substantial fluctuations in precipitation dynamics observed across the years. In contrast, the Ny-{\AA}lesund dataset exhibits a slightly higher mean weekly precipitation of 8.98 mm with a significant variability, spanning from 0 mm to 179.1 mm. The coefficient of variation reaches 158.28\%, reflecting extremely pronounced fluctuations and highlighting the erratic nature of precipitation dynamics in Ny-{\AA}lesund compared to Bj{\o}rn{\o}ya. Furthermore, to determine the structural characteristics of the precipitation dynamics in the Arctic, we examine the following global features:
\begin{itemize} 
    \item \textbf{Stationarity:} We assess whether the statistical properties of the series, such as mean and variance, remain constant over time. The Kwiatkowski Phillips Schmidt Shin (KPSS) test is applied using the \textit{kpss.test} function from the \texttt{tseries} package in $\mathbf{R}$ to evaluate stationarity.

    \item \textbf{Linearity:} Determining whether a time series exhibits linear or nonlinear dynamics is crucial, as it directly influences the selection of appropriate forecasting models. To assess this, we apply Terasvirta’s neural network test using the \texttt{nonlinearTseries} package in $\mathbf{R}$.

    \item \textbf{Seasonality:} To identify recurring temporal patterns, we use the combined seasonality test proposed by Ollech and Webel, implemented via the \textit{isSeasonal} function from the \texttt{seastests} package in $\mathbf{R}$.

    \item \textbf{Long-range dependency:} The presence of self-similar behavior or persistence in the time series is examined by estimating the Hurst exponent using the \textit{hurstexp} function from the \texttt{pracma} package in $\mathbf{R}$.

    \item \textbf{Normality:} Assessing normality is essential, as many statistical models assume Gaussian behavior and violations of this assumption can significantly influence model performance. We apply the Anderson–Darling test using the \textit{ad.test} function from the \texttt{nortest} package in $\mathbf{R}$ to evaluate the normality of the series. 
\end{itemize}
The results of the statistical tests indicate that the weekly precipitation levels in both the Bear Island and Ny-{\AA}lesund regions exhibit stationary behavior and long-term dependence, as reflected by Hurst exponent values of 0.503 and 0.584, respectively. The Bj{\o}rn{\o}ya dataset, however, significantly departs from the assumptions of linearity and normality, exhibiting nonlinear dynamics and a right-skewed distribution with heavy tails, largely driven by extreme precipitation events. Seasonal patterns are also evident in Bj{\o}rn{\o}ya, highlighting periodic components in its precipitation dynamics. On the other hand, Ny-{\AA}lesund displays a predominantly linear behavior with a right-skewed, heavy-tailed distribution, and its precipitation dynamics do not exhibit pronounced seasonal fluctuations. To further examine the temporal structure of the data, we visualize the training dataset along with its autocorrelation function (ACF) and partial autocorrelation function (PACF) in Figure \ref{Fig_ACF_PACF_Plot}. The precipitation time series for Bj{\o}rn{\o}ya exhibits several distinctive patterns. The temporal dynamics are characterized by several high precipitation events with long periods of low or no rainfall, resulting in a right-skewed distribution with heavy tails, which reflects the occurrence of extreme precipitation. Seasonal fluctuations are evident, with recurring peaks suggesting a periodic component in the weekly precipitation dynamics. The ACF plot of Bj{\o}rn{\o}ya decays gradually over multiple lags, indicating the presence of long-term dependence and persistent temporal correlations, while the PACF shows significant spikes at short lags, reflecting strong short-term autocorrelation. Together, these patterns indicate that precipitation in Bj{\o}rn{\o}ya is both highly variable and influenced by historical observations, with underlying seasonality and nonlinear dynamics. In the case of the Ny-{\AA}lesund, precipitation series displays a different behavior. While the distribution is also right-skewed with heavy tails, extreme events are more irregular. The series shows a linear trajectory compared to Bj{\o}rn{\o}ya, with fewer abrupt fluctuations. It's ACF decays rapidly, suggesting weaker long-term dependence, and the PACF exhibits only a few significant lags, indicating that past precipitation has limited influence on future values. Furthermore, Ny-{\AA}lesund lacks clear seasonal cycles, implying that the precipitation dynamics are dominated by short-term variability rather than periodic patterns. These differences highlight that, although both Arctic regions experience extreme events, the temporal structure, seasonality, and persistence of precipitation differ markedly, which has important implications for evaluating the generalizability of the forecasting techniques.

\subsection{Forecasting Methods}
Accurate precipitation forecasts are essential for effective planning and decision-making across various sectors. However, precipitation forecasting remains an inherently challenging task due to its complex, stochastic, and nonlinear nature. The influence of numerous atmospheric and climatic variables introduces significant variability, making it difficult for conventional statistical models to capture the underlying dynamics with sufficient accuracy. To address these complexities, this study evaluates a diverse set of forecasting techniques drawn from both machine learning and deep learning paradigms. Each model is assessed under two settings: one based solely on historical precipitation data and another incorporating key exogenous climatic drivers, such as temperature, relative humidity, cloud cover, and air pressure, which have been identified to influence precipitation behavior causally (in Section \ref{sec:Causality}). This evaluation strategy provides a more comprehensive understanding of how auxiliary climate information can enhance forecast performance. The models considered in the analysis include modified linear forecasters, tree-based boosting algorithms, recurrent neural network architectures, deep neural networks, and attention-based frameworks. The selection is motivated not only by their established use in forecasting literature but also by their capacity to model the dynamic behavior of precipitation across different timescales and under varying conditions. In this section, we provide an overview of each of the forecasting techniques used in our analysis.

\begin{itemize}
    
    \item \emph{Decomposition-based Linear} (DLinear) model is a modified linear time series forecasting technique that explicitly decomposes the input time series into two components and models them separately \cite{zeng2023transformers}. The architecture employs a moving average kernel to extract the trend component, while the remainder serves as the seasonal component. Each of the components are individually modeled using a linear layer and the aggregated output from these linear layers provides the final prediction. By isolating long-term trends and periodic variations, DLinear avoids the complexity of deeper architectures while retaining robust forecasting performance. 

    \item \emph{Normalization-based Linear} (NLinear) is a variant of the DLinear framework designed to address distributional shifts in time series datasets \cite{zeng2023transformers}. Unlike DLinear, which decomposes the time series into trend and seasonal components, NLinear adopts a normalization-based strategy to stabilize the data. Specifically, the input sequence is normalized by subtracting its last observed value, reducing variability due to level shifts. The normalized sequence is then modeled using a linear layer, after which the subtracted component is added back to produce the final forecast. This simple normalization approach helps the model generalize better by reducing the impact of distributional shifts between the training and test datasets, enhancing its robustness.

    \item \emph{Random Forest} is a widely used supervised learning algorithm that has been effectively adapted for time series forecasting tasks \cite{breiman2001random}. In this framework, an ensemble of decision trees is constructed during training, with each tree trained on a distinct blocked bootstrap sample of the time series. The blocked bootstrap approach, where contiguous segments of the time series are randomly sampled and concatenated, helps preserve temporal dependencies and structure within the data \cite{goehry2023random}. The final forecast is obtained by averaging the predictions of all individual trees, which reduces variance and enhances the model's robustness. This ensemble-based design enables random forest to capture complex nonlinear relationships and temporal dependencies among historical observations, making it a strong baseline for comparison in time series forecasting studies.

    \item \emph{Long Short-term Memory} (LSTM) networks are an advanced variant of classical recurrent neural networks (RNNs) designed to capture long-term dependencies in sequential data \cite{hochreiter1997long}. They address the vanishing and exploding gradient problems commonly faced by standard RNNs through a carefully designed gating mechanism. The core innovation of LSTMs lies in their memory cell structure, which is controlled by three gates: the \textit{forget gate}, \textit{input gate}, and \textit{output gate}. These gates regulate the flow of information within the cell state, which acts as long-term memory, and the hidden state, representing short-term dynamics. This selective memory mechanism enables LSTMs to retain relevant past information over extended time horizons, making them particularly effective for time series forecasting tasks where distant historical observations influence future outcomes.
    \item \emph{Neural Basis Expansion Analysis for Time Series} (NBeats) model is a deep learning architecture designed for time series forecasting tasks \cite{oreshkin2019n}. It is built using a fully connected feedforward network organized into a stack of sequential blocks that operate recursively. Each block encompasses two layers: one layer attempts to reconstruct the input signal and produce a preliminary forecast, while the other layer refines the forecasts by utilizing an error correction approach. This recursive architecture enables NBeats to effectively capture both long-term trends and short-term fluctuations in time series data, resulting in better forecasting performance across a wide range of applied domains. 
    
    \item \emph{Neural Hierarchical Interpolation for Time Series} (NHiTS) model extends the theoretical foundations of NBeats by introducing a multi-scale hierarchical structure \cite{challu2023nhits}. Its block-based framework begins by downsampling the input sequence to extract low-frequency temporal features, which are then refined through interpolation-based upsampling. This design allows the model to learn patterns across multiple temporal scales, from low-frequency trends to fine-grained fluctuations. Residual connections across blocks help iteratively reduce forecasting errors, enhancing performance over long horizons. Through the combination of multi-resolution feature extraction and efficient interpolation, NHiTS provides a robust and scalable solution for long-range time series forecasting.

    \item \emph{Transformers} represent a class of state-of-the-art deep learning models that utilize self-attention mechanisms to capture long-range dependencies in time series data \cite{wu2020deep}. In contrast to RNNs, Transformers process entire input sequences in parallel, significantly enhancing computational efficiency and scalability. The multi-head attention mechanism allows the model to handle multiple time steps simultaneously, enabling it to learn complex interactions and temporal dependencies within the historical data. This architecture has proven particularly effective in modeling non-linear and multi-scale dynamics in time series forecasting tasks.

    \item \emph{Time-series Dense Encoder} (TiDE) is a multi-layer perceptron (MLP)-based encoder-decoder architecture developed for long-term time series forecasting \cite{das2023long}. The encoder transforms historical time series data into a dense latent representation through feature projections and fully connected encoding layers. The decoder then maps this latent representation into future forecasts. Both the encoder and decoder incorporate residual blocks with multiple hidden layers and skip connections, enabling the model to capture complex temporal patterns effectively. The architectural design of TiDE allows it to address key limitations of existing forecasting models, such as the suboptimal performance of Transformer-based architectures in long-range settings and the inability of linear models to capture nonlinear dependencies, offering a robust and scalable alternative for time series forecasting tasks.

    \item \emph{eXtreme Gradient Boosting} (XGBoost) is an ensemble learning technique that improves gradient-boosted decision trees via parallel computing and better optimization. For time series forecasting, the problem is restructured as a supervised learning setup where input features are the lagged values of the time series, while the output labels are the current values of the series. Mathematically, the XGBoost model builds an ensemble of $\mathcal{E}$ additive regression trees, where the prediction at time $t$ is given by:
    $$
    \hat{y}_t=\sum_{e=1}^{\mathcal{E}} f_e\left(\underline{y}_{t-1}\right), 
    $$
    where $\underline{y}_{t-1} = \{y_{t-1}, y_{t-2}, \ldots, y_{t-p}\}$ represents the input feature vector composed of $p$ lagged values at time $t$ and $f_e$ denotes the $e^{th}$ regression tree. The model is trained by minimizing a regularized objective function:
    $$
    \operatorname{obj}(\theta) = \sum_{t} \ell(y_t, \hat{y}_t) + \sum_{e = 1}^{\mathcal{E}} \Omega(f_e),
    $$
    where $\ell(y_t, \hat{y}_t)$ is a differentiable loss function (such as the mean squared loss) and $\Omega(f_e)$ is a regularization term that penalizes model complexity, such that 
    $$
    \Omega(f_e) = \lambda \;T_e + \frac{1}{2} \gamma \|\omega_e\|^2
    $$
    with $T_e$ being the number of leaves in the $e$-th tree, $\omega_e$ the vector of leaf weights, and $(\gamma, \lambda)$ are the regularization hyperparameters. By optimizing this objective iteratively, XGBoost sequentially adds trees to correct the residuals of the previous ensemble, allowing it to capture complex nonlinear and interaction effects among historical observations. This makes it particularly suitable for modeling the nonlinear and multi-scale dependencies often observed in time series data.

    \item[] To further evaluate the role of auxiliary climatic information, we examine the exogenous variants of the aforementioned models, where historical climatic variables are incorporated alongside lagged precipitation data as model inputs. By including these causal covariates, we extend each forecasting framework to its exogenous form, namely, DLinearX, NLinearX, Random ForestX, LSTMX, NBeatsX, NHiTSX, TransformerX, TiDEX, and XGBoostX. The following section presents a comprehensive evaluation of both univariate (precipitation-only) and multivariate (climate-aware) variants of these models, thereby assessing their effectiveness in capturing precipitation dynamics.
\end{itemize}

\subsection{Performance Indicators}
In our analysis, we assess the performance of various forecasting methods using both scale-dependent and relative error metrics. For scale-dependent evaluation, we employ the Root Mean Squared Error (RMSE) and the Mean Absolute Error (MAE), which quantify forecast performance in the same units as the original data. To facilitate comparison across series with differing scales, we further incorporate relative error measures, namely, the symmetric Mean Absolute Percentage Error (sMAPE) and the Mean Absolute Ranged Relative Error (MARRE) \cite{hyndman2018forecasting, panja2024stgcn}. The corresponding mathematical formulations are as follows:
\[ \text{ RMSE} = \sqrt{\frac{1}{h}\sum_{t =1}^{h} (y_t - \hat{y}_t)^2}; \; \;
\text{ MAE} = \frac{1}{h}\sum_{t =1}^{h} |y_t - \hat{y}_t|; \;
\;\]
\[
\text{ sMAPE} = \frac{1}{h} \sum_{t=1}^h \frac{2|\hat{y}_t - y_t|}{|\hat{y}_t|+ |y_t|} \times 100 \% \; \text{ and } \] 
\[\text{ MARRE} = \frac{1}{h} \sum_{t=1}^h\left|\frac{y_t-\hat{y}_t}{\max _t y_t-\min _t y_t}\right| \times 100 \%;
\]
where $y_t$ represents the ground truth observations, $\hat{y}_t$ is the corresponding forecast at time $t$, and $h$ is the forecast horizon. By definition, lower values of these metrics indicate better forecasting performance, with the minimum value conventionally signifying the most accurate model.

\subsection{Forecasting Performance Evaluation}

In this study, we focus on one-week-ahead forecasting of precipitation during the testing period. To ensure a consistent and fair evaluation across all forecasting methods, we use three lagged observations of precipitation as inputs in the univariate setting. For the multivariate setup, we include three historical observations, each of precipitation and the associated causal climatic variables: temperature, relative humidity, cloud cover, and air pressure. {For the deep learning models (e.g., LSTM), training was performed using the Adam optimizer with an initial feature map size of 25, a batch size of 32, and a maximum of 100 epochs. In the case of Transformers, we used the multi-head attention mechanism with the number of heads set to 4. Notably, the same hyperparameter configuration was applied across all forecasting frameworks to ensure a consistent and fair evaluation.

\begin{table*}[!ht]
    \centering
    \caption{ Performance measures for forecasting precipitation dynamics in Bear Island and Ny-{\AA}lesund regions. Best results are \textbf{highlighted}.}
     \begin{tabular}{|c|cccc|cccc|}
        \hline
         \multirow{2}{*}{Model}   & \multicolumn{4}{c|}{Bear Island} & \multicolumn{4}{c|}{Ny-{\AA}lesund} \\ 
        & RMSE & MAE & SMAPE (\%) & MARRE (\%) & RMSE & MAE & SMAPE (\%) & MARRE (\%) \\ \hline
        Dlinear & 9.2083 & 6.5774 & 82.3248 & 15.9360 & 12.3293 & 8.5849 & 112.1141 & 17.2388 \\
        DlinearX & 9.1498 & 6.5028 & 80.2902 & 12.9790 & 6.5259 & 4.9287 & 40.8597 & 15.7939 \\
        Nlinear & 9.5426 & 6.5529 & 84.2671 & 14.4805 & 14.0386 & 9.8645 & 119.4452 & 19.8083 \\ 
        NlinearX & 9.1989 & 6.5065 & 83.9147 & 13.1977 & 7.0373 & 5.1185 & 42.6378 & 16.7101 \\
        Random Forest & 11.0555 & 7.4574 & 89.0771 & 15.1266 & 13.9149 & 9.7275 & 112.0707 & 19.5332 \\ 
        Random ForestX & 9.7244 & 7.2028 & 88.5179 & 14.6101 & 6.4867 & 4.7851 & 39.8874 & 15.1478 \\ 
        LSTM & 9.2738 & 7.6423 & 85.3241 & 13.7731 & 12.5448 & 9.0677 & 114.1467 & 18.2082 \\ 
        LSTMX & 9.1765 & 6.7144 & 85.1463 & 13.6194 & 10.7815 & 8.6766 & 98.6739 & 16.1517 \\ 
        NBeats & 9.9729 & 7.6811 & 85.4816 & 14.6519 & 12.2391 & 8.7639 & 112.9191 & 17.5982 \\
        NBeatsX & 9.5403 & 6.4176 & 79.4301 & 12.9891 & 6.6219 & 5.1224 & 41.5056 & 16.4685 \\
        NHiTS & 9.6246 & 6.5989 & 84.9469 & 16.3851 & 12.3368 & 8.9488 & 113.2464 & 17.9695 \\ 
        NHiTSX & 9.5771 & 7.8371 & 92.3137 & 15.8969 & 6.5734 & 4.7219 & 44.0331 & 17.0936  \\ 
        Transformer & 9.2069 & 6.9758 & 86.8667 & 13.9412 & 12.2018 & 8.7899 & 110.7074 & 17.6503 \\ 
        TransformerX & 9.1017 & 6.8107 & 85.8486 & 13.8148 & 7.3112 & 5.6875 & 51.1431 & 19.2182 \\ 
        TiDE & 9.9933 & 7.5089 & 87.6950 & 14.2026 & 12.4708 & 9.3200 & 113.2383 & 18.7148 \\ 
        TiDEX & 9.1603 & 6.9053 & 86.9262 & 14.0068 & 6.8793 & 4.7630 & 45.9845 & 17.0410 \\         XGBoost & 11.9268 & 7.5793 & 86.2722 & 15.3739 & 15.1871 & 9.8921 & 115.4925 & 19.8636 \\ 
        XGBoostX & \textbf{9.0691} & \textbf{6.4053} & \textbf{78.8194 }& \textbf{12.9725} & \textbf{6.4712} & \textbf{4.6711} & \textbf{39.6991} & \textbf{15.1524} \\  \hline
    \end{tabular}
    \label{Table_Performance_Comp}
\end{table*}

Table \ref{Table_Performance_Comp} summarizes the forecasting performance of various models for precipitation dynamics in the Bj{\o}rn{\o}ya and Ny-{\AA}lesund regions, evaluated using key error metrics. The results indicate that the XGBoostX consistently delivers the best performance across all four error metrics for both regions.} This highlights its ability to capture complex, nonlinear relationships between precipitation and the key climatic drivers. The inclusion of causal climatic variables significantly enhances its modeling capability, allowing XGBoostX to capture the underlying precipitation dynamics more accurately than its univariate counterpart. While the relative ranking of the forecasting techniques remains broadly consistent, notable regional differences emerge. In Bear Island, where precipitation is comparatively less volatile, XGBoostX achieves particularly lower SMAPE and MARRE values, outperforming competitive neural architectures like NHiTSX and NBeatsX by a significant margin. In Ny-{\AA}lesund, which exhibits higher precipitation variability and stronger fluctuations, the advantage of XGBoostX is even more pronounced: it delivers the lowest RMSE and MAE, while complex deep learning models such as LSTM, TransformerX, and TiDEX struggle with generalization. This is consistent with the low-data regime problems of deep learning-based architectures, which require larger historical observations and more stable dynamics for effective training. In contrast, XGBoostX is less dependent on large volumes of data and maintains stable performance through its built-in regularization and pruning techniques. While Random ForestX also adopts an ensemble learning strategy, similar to XGBoostX, its reliance on bagging, rather than boosting, limits its ability to iteratively improve and correct prediction errors. As a result, it is less effective at capturing complex interactions and long-range dependencies in the data. Neural network-based forecasting models like NBeatsX, NHiTSX, and LSTMX are capable of modeling long-term temporal patterns, but their performance is negatively impacted by the presence of heterogeneous covariates. These models often require careful tuning and larger training sets to perform optimally, whereas XGBoostX offers more consistent results across different input configurations. Furthermore, modified linear models such as DLinearX and NLinearX depend on structural assumptions like additive decomposition and distributional shift, which are often violated in precipitation datasets that exhibit irregular fluctuations and nonlinear behavior, as discussed in Section \ref{Sec_Global_Features}. Their limited flexibility makes them less competitive in this context. Overall, XGBoostX provides the most robust and effective architecture for precipitation forecasting in this study. 

\begin{figure}
    \centering
    \includegraphics[width=\linewidth]{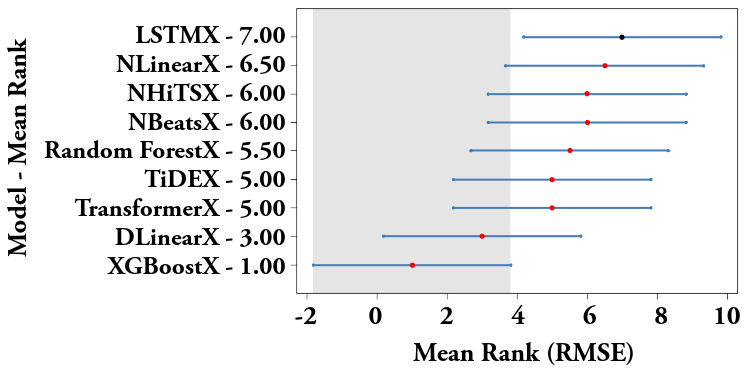}
    \caption{Visualization of the multiple comparisons with the best MCB test in terms of RMSE metric. The horizontal axis depicts the average ranks, while the vertical axis represents the forecasting technique, such that {XGBoostX-1.00} indicates the average rank of the XGBoostX framework based on the RMSE metric is 1.00, and similar to others.}
    \label{fig_MCB}
\end{figure}

Furthermore, to validate the statistical significance of the performance improvement in the XGBoostX framework, we conduct Multiple Comparison with the Best (MCB) test \cite{koning2005m3}. This distribution-free test procedure ranks the competitive models based on their performance indicators for different datasets and computes the average rank with their corresponding critical distances. The model with the least average rank is considered the `best' performing framework, and its critical distance serves as the reference value for this test. The MCB test applied to multivariate forecasting techniques using the RMSE metric, as shown in Figure \ref{fig_MCB}, highlights that the XGBoostX framework attains the minimum rank and is considered the `best' performing approach, followed by DLinearX and TransformerX. This superior performance of the XGBoostX framework is attributed to several factors inherent to both the climatic data characteristics and the model architecture. First, precipitation dynamics in Bear Island are driven by complex nonlinear interactions with temperature, humidity, cloud cover, and pressure, which are effectively captured by the gradient-boosted decision trees of the XGBoostX framework. Second, the limited size and noisy nature of Arctic datasets favor models that are robust in small-data regimes; XGBoostX is particularly well-suited in these scenarios due to its built-in regularization and ensemble averaging, whereas deep neural architectures are more prone to overfitting. Third, the tree-based structure enables adaptive partitioning of the lagged climatic variables, allowing the model to selectively emphasize the most influential drivers in different temporal contexts, thereby improving both accuracy and interpretability. These inherent properties make the XGBoostX model more suitable than the classical statistical models and deep learning approaches in forecasting precipitation dynamics of Arctic regions.

\subsection{Uncertainty Quantification}
\begin{figure*}
    \centering
    \includegraphics[width=\linewidth]{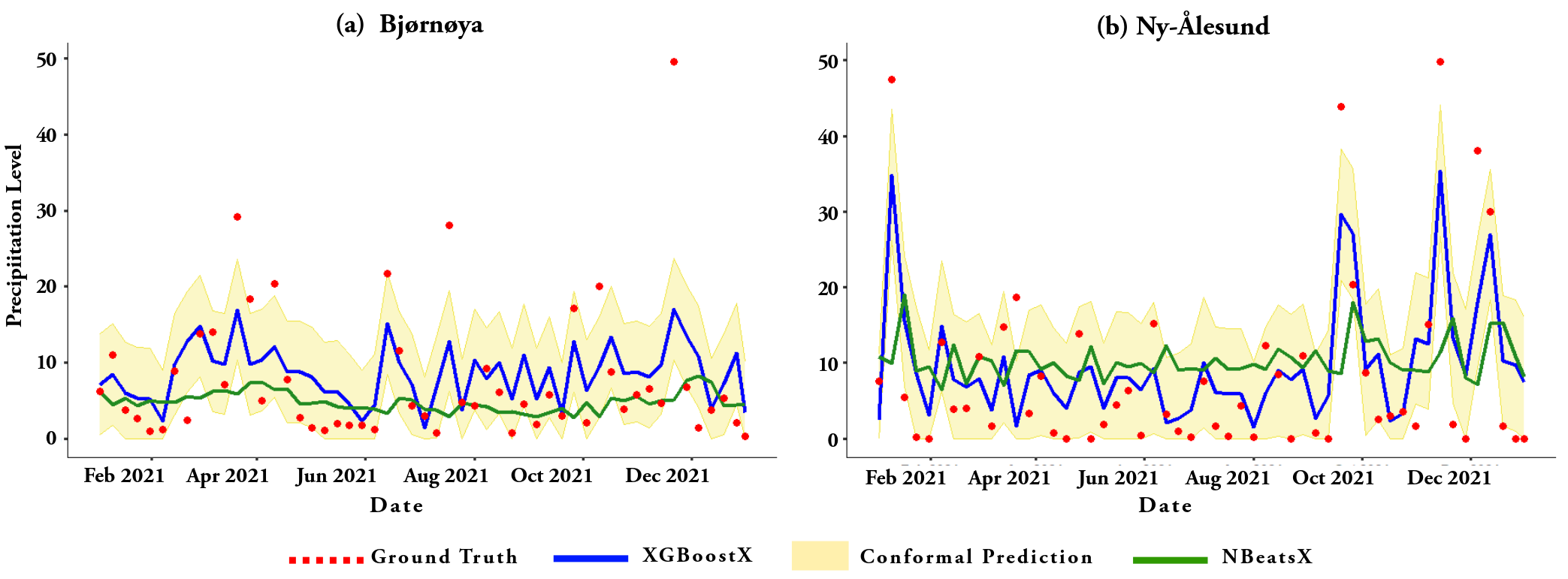}
    \caption{The plot visualizes the ground truth (red points) precipitation dynamics observed in (a) Bj{\o}rn{\o}ya and (b) Ny-{\AA}lesund regions during 2021, along with the forecasts of the XGBoostX model (blue), forecasts of the NBeatsX model (green), and the probabilistic band (based on the conformal prediction approach) of the XGBoostX framework (yellow shaded).}
    \label{Fig_CP_Plot}
\end{figure*}

In addition to the point forecasts, we employed the conformal prediction approach to assess the uncertainties associated with the XGBoostX model's precipitation forecasts. The model-agnostic conformal prediction approach provides a non-parametric procedure to convert point estimates of precipitation dynamics into suitable prediction intervals \cite{vovk2005algorithmic}. Given the input sequence of precipitation dynamics $\{y_t\}$ and climatic drivers $\mathrm{X}_t=\left\{x_{1, t}, x_{2, t}, \ldots, x_{N, t}\right\}$, this approach fits an uncertainty model ($\mathrm{UM}$) on the lagged observations $\left\{\underline{y}_{t-1}, \underline{\mathrm{X}}_{t-1}\right\}$ to generate a notion of uncertainty. Thus, the conformal score ($\mathrm{CS}_t$) can be computed as 
$$
    \mathrm{CS}_t = \frac{\left|y_t - \mathrm{XGBoostX}\left(\underline{y}_{t-1}, \underline{\mathrm{X}}_{t-1}\right)\right|}{\mathrm{UM}\left(\underline{y}_{t-1}, \underline{\mathrm{X}}_{t-1}\right)}.
$$
By leveraging the sequential patterns of the input series, the conformal quantile $\left(\mathrm{CQ}_t\right)$ can be computed using a weighted conformal method with an $\alpha$-sized window $\Gamma_t = \mathbf{1}\left(\tilde{t} \geq t - \alpha\right), \forall \; \tilde{t} < t$ as 
\begin{equation*}
    \mathrm{CQ}_t = \inf\left\{\Delta: \frac{1}{\min\left(\alpha, \tilde{t} -1\right) + 1} \sum_{\tilde{t} = 1}^{t-1} \mathrm{CS}_{\tilde{t}} \Gamma_{\tilde{t}} \geq 1- \delta \right\}.
\end{equation*}
Thus, the $100*\left(1- \delta\right)\%$ conformal prediction interval based on these weighted quantiles is given by:
\begin{equation*}
    \mathrm{XGBoostX}\left(\underline{y}_{t-1}, \underline{\mathrm{X}}_{t-1}\right) \pm \text{CQ}_t \; \mathrm{UM}\left(\underline{y}_{t-1}, \underline{\mathrm{X}}_{t-1}\right).
\end{equation*}
In this analysis, we derive 90\% conformal prediction intervals for the precipitation forecasts of Bj{\o}rn{\o}ya and Ny-{\AA}lesund regions produced by the XGBoostX model and present the results in Figure \ref{Fig_CP_Plot}. To ensure the validity of these intervals and prevent data leakage, residuals are computed using a separate calibration (validation) set distinct from the test data. Alongside the prediction intervals, we also display the point forecasts from both the XGBoostX and NBeatsX models during the test period, along with the corresponding ground-truth precipitation values. As evident from the plot, the XGBoostX framework captures the essential features of precipitation dynamics in Bear Island and Ny-{\AA}lesund. In both regions, the model tracks the irregular variability marked by alternating dry spells and sudden rainfall spikes, which are characteristic of Arctic precipitation. In Bj{\o}rn{\o}ya, the forecasts closely follow the moderate but frequent fluctuations, while in Ny-{\AA}lesund the model effectively captures the timing and magnitude of sharper extremes. The widening of the conformal prediction bands during high-variability periods reflects the increased uncertainty associated with extreme events, consistent with the stochastic nature of Arctic precipitation. Compared to the smoother NBeatsX forecasts, XGBoostX is more responsive to sudden shifts and better represents the heavy-tailed distribution of precipitation, underscoring its robustness in modeling both persistent variability and rare extremes. Overall, the XGBoostX framework, augmented with conformal prediction, provides a reliable and probabilistic representation of precipitation dynamics, enhancing its value for disaster preparedness and risk management. However, the model fails to accurately predict some of the rare but high-impact extreme precipitation events that are inherently underrepresented in historical data and are not adequately addressed by residual-based uncertainty quantification. Forecasting such events requires specialized tail modeling techniques, such as Extreme Value Theory (EVT), which can be integrated into data-driven architectures to better capture both the temporal evolution and the statistical properties of rare precipitation extremes. Incorporating EVT into the current framework would strengthen its capacity for extreme precipitation forecasting and improve its robustness in climate-sensitive Arctic environments.

\section{\label{conclusion} Conclusion and Discussion}
Precipitation in the Arctic is a critical climate concern, with far-reaching implications for sea level rise, glacier melt, ocean circulation, and ecosystem disruption. Bear Island and Ny-{\AA}lesund, situated in this rapidly changing environment, provide a distinct setting for studying such events. Despite the availability of observational data, accurately forecasting precipitation dynamics in the Arctic regions remains a challenging task due to complex, nonlinear interactions among atmospheric drivers such as temperature, humidity, cloud cover, and air pressure. This study addresses these challenges through a data-driven framework that models these interactions and enables probabilistic forecasting by incorporating both historical precipitation and exogenous climatic variables. 

We propose a comprehensive approach by integrating individual-level and joint-level causality analyses with data-driven forecasting and uncertainty quantification techniques. Our approach addresses critical challenges of sparse data, strong atmospheric variability, and limited operational predictability of Arctic precipitation dynamics. In the integrated framework, wavelet coherence analysis offers valuable insights into the time-frequency relationships between precipitation and key atmospheric drivers, while the SURD approach provides a deeper understanding of how these variables jointly influence precipitation dynamics. These insights are incorporated into data-driven forecasting models to improve forecast accuracy and interpretability. Furthermore, to account for uncertainty in future precipitation dynamics, conformal prediction intervals are employed. These probabilistic forecasts enhance the practical utility of our approach for guiding early warning systems in climate-sensitive Arctic regions. Empirical results show that tree-based ensemble models, particularly XGBoost with exogenous climatic drivers, outperform deep neural architectures in terms of forecast performance, especially in data-scarce settings like Bear Island and  Ny-{\AA}lesund. Overall, the integrated framework presented in this study provides a technological solution for forecasting the temporal evolution of precipitation dynamics in the Arctic and offers a reliable tool to support informed risk mitigation initiatives. 

Despite the practical utility of the XGBoostX framework, it fails to adequately capture the rare and high-impact extreme precipitation events that can result in severe consequences in the Arctic environment. To address this key limitation of the current work, future studies can focus on integrating Extreme Value Theory with data-driven architecture to better model the tail behavior of precipitation dynamics in the Arctic. This combined approach will enhance the modeling capability of the framework to generate risk-aware precipitation forecasts. Further studies can also focus on extending this framework by incorporating spatial dependencies across neighboring Arctic regions and generating accurate spatiotemporal forecasts for precipitation dynamics in the broader Arctic marine environment. Another potential direction is the integration of physical knowledge of the governing systems into the forecasting models, which could enhance their generalizability and robustness, particularly under data-scarce conditions of polar regions.

\section*{Data Availability Statement}
The original data is collected from publicly accessible sources \cite{datasource}. All codes for performing preliminary analysis, identifying causal variables, and implementing data-driven forecasting models, alongside the processed data, are made available on Zenodo \cite{codesource}.

\section*{Acknowledgment}
M. Panja would like to acknowledge Professor Alejandro Tejedor of the University of Zaragoza, Zaragoza, Spain, for providing insightful suggestions on the first version of this manuscript, which significantly improved the paper. D. Das thanks the University Grants Commission (UGC), India, for providing financial support (NTA Ref. No. 211610125989). D. Ghosh was supported by the Science and Engineering Research Board (SERB), Government of India (Project No. CRG/2021/005894). The authors would also like to thank the editor and the two learned reviewers for their constructive discussions and valuable comments.

\section*{Dedication}
Professor Ljupco Kocarev is a renowned researcher in nonlinear dynamics, network science, and machine learning.  In the mid-1990s, he introduced the concept of chaotic synchronization for secure communication systems~\cite{kocarev1995general, kocarev2005complex}.  He went on to expand our understanding of complex networks and popular phenomena~\cite{kocarev2013consensus}, and more recently, he assisted in connecting nonlinear dynamics with modern machine learning methodologies~\cite{tang2020introduction}.  In this manuscript, we present a probabilistic machine learning framework for modeling and forecasting precipitation extremes in Arctic environments, which is closely coincides with his works on nonlinear dynamics, causal inference, and the integration of machine learning into complex systems research.

\section*{AUTHOR DECLARATIONS}
\section*{Conflict of Interest}
The authors have no conflicts to disclose.

\section*{Author Contributions}
{\bf Madhurima Panja}: Conceptualization (equal); Data curation (equal); Formal analysis (equal); Investigation (equal); Methodology
(equal); Software (equal); Validation (equal); Visualization (equal); Writing – original draft (equal). 

{\bf Dhiman Das}: Conceptualization (equal); Data curation (equal); Formal analysis (equal); Investigation (equal); Methodology
(equal); Software (equal); Validation (equal); Visualization (equal); Writing – original draft (equal). {\bf Tanujit Chakraborty}: Conceptualization (equal); Supervision (equal); Visualization
(equal); Writing – review and editing (equal). {\bf Arnob Ray}: Conceptualization (equal); Data curation (equal); Formal analysis (equal); Visualization (equal); Writing – review and editing (equal).
 {\bf R. Athulya}: Conceptualization (equal); Data curation (equal); Formal analysis (equal); Visualization (equal); Writing – review and editing (equal).
{\bf Chittaranjan Hens}: Conceptualization (equal); Supervision (equal); Visualization
(equal); Writing – review and editing (equal). 
{\bf Syamal K Dana}: Conceptualization (equal); Supervision (equal); Visualization
(equal); Writing – review and editing (equal).
{\bf Nuncio Murukesh}: Conceptualization (equal); Supervision (equal); Visualization
(equal); Writing – review and editing (equal).
{\bf Dibakar Ghosh}: Conceptualization (equal); Supervision (equal); Visualization
(equal); Writing – review and editing (equal).  

\section*{References}
\bibliography{precipitation_ref} 

\begin{thebibliography}{67}%
\makeatletter
\providecommand \@ifxundefined [1]{%
 \@ifx{#1\undefined}
}%
\providecommand \@ifnum [1]{%
 \ifnum #1\expandafter \@firstoftwo
 \else \expandafter \@secondoftwo
 \fi
}%
\providecommand \@ifx [1]{%
 \ifx #1\expandafter \@firstoftwo
 \else \expandafter \@secondoftwo
 \fi
}%
\providecommand \natexlab [1]{#1}%
\providecommand \enquote  [1]{``#1''}%
\providecommand \bibnamefont  [1]{#1}%
\providecommand \bibfnamefont [1]{#1}%
\providecommand \citenamefont [1]{#1}%
\providecommand \href@noop [0]{\@secondoftwo}%
\providecommand \href [0]{\begingroup \@sanitize@url \@href}%
\providecommand \@href[1]{\@@startlink{#1}\@@href}%
\providecommand \@@href[1]{\endgroup#1\@@endlink}%
\providecommand \@sanitize@url [0]{\catcode `\\12\catcode `\$12\catcode `\&12\catcode `\#12\catcode `\^12\catcode `\_12\catcode `\%12\relax}%
\providecommand \@@startlink[1]{}%
\providecommand \@@endlink[0]{}%
\providecommand \url  [0]{\begingroup\@sanitize@url \@url }%
\providecommand \@url [1]{\endgroup\@href {#1}{\urlprefix }}%
\providecommand \urlprefix  [0]{URL }%
\providecommand \Eprint [0]{\href }%
\providecommand \doibase [0]{http://dx.doi.org/}%
\providecommand \selectlanguage [0]{\@gobble}%
\providecommand \bibinfo  [0]{\@secondoftwo}%
\providecommand \bibfield  [0]{\@secondoftwo}%
\providecommand \translation [1]{[#1]}%
\providecommand \BibitemOpen [0]{}%
\providecommand \bibitemStop [0]{}%
\providecommand \bibitemNoStop [0]{.\EOS\space}%
\providecommand \EOS [0]{\spacefactor3000\relax}%
\providecommand \BibitemShut  [1]{\csname bibitem#1\endcsname}%
\let\auto@bib@innerbib\@empty
\bibitem [{\citenamefont {Kangalawe}\ \emph {et~al.}(2017)\citenamefont {Kangalawe}, \citenamefont {Mung’ong’o}, \citenamefont {Mwakaje}, \citenamefont {Kalumanga},\ and\ \citenamefont {Yanda}}]{kangalawe2017climate}%
  \BibitemOpen
  \bibfield  {author} {\bibinfo {author} {\bibfnamefont {R.~Y.}\ \bibnamefont {Kangalawe}}, \bibinfo {author} {\bibfnamefont {C.~G.}\ \bibnamefont {Mung’ong’o}}, \bibinfo {author} {\bibfnamefont {A.~G.}\ \bibnamefont {Mwakaje}}, \bibinfo {author} {\bibfnamefont {E.}~\bibnamefont {Kalumanga}}, \ and\ \bibinfo {author} {\bibfnamefont {P.~Z.}\ \bibnamefont {Yanda}},\ }\href@noop {} {\bibfield  {journal} {\bibinfo  {journal} {Climate and Development}\ }\textbf {\bibinfo {volume} {9}},\ \bibinfo {pages} {202} (\bibinfo {year} {2017})}\BibitemShut {NoStop}%
\bibitem [{\citenamefont {Huq}\ \emph {et~al.}(2004)\citenamefont {Huq}, \citenamefont {Reid}, \citenamefont {Konate}, \citenamefont {Rahman}, \citenamefont {Sokona},\ and\ \citenamefont {Crick}}]{huq2004mainstreaming}%
  \BibitemOpen
  \bibfield  {author} {\bibinfo {author} {\bibfnamefont {S.}~\bibnamefont {Huq}}, \bibinfo {author} {\bibfnamefont {H.}~\bibnamefont {Reid}}, \bibinfo {author} {\bibfnamefont {M.}~\bibnamefont {Konate}}, \bibinfo {author} {\bibfnamefont {A.}~\bibnamefont {Rahman}}, \bibinfo {author} {\bibfnamefont {Y.}~\bibnamefont {Sokona}}, \ and\ \bibinfo {author} {\bibfnamefont {F.}~\bibnamefont {Crick}},\ }\href@noop {} {\bibfield  {journal} {\bibinfo  {journal} {Climate Policy}\ }\textbf {\bibinfo {volume} {4}},\ \bibinfo {pages} {25} (\bibinfo {year} {2004})}\BibitemShut {NoStop}%
\bibitem [{\citenamefont {Vihma}\ \emph {et~al.}(2016)\citenamefont {Vihma}, \citenamefont {Screen}, \citenamefont {Tjernstr{\"o}m}, \citenamefont {Newton}, \citenamefont {Zhang}, \citenamefont {Popova}, \citenamefont {Deser}, \citenamefont {Holland},\ and\ \citenamefont {Prowse}}]{vihma2016atmospheric}%
  \BibitemOpen
  \bibfield  {author} {\bibinfo {author} {\bibfnamefont {T.}~\bibnamefont {Vihma}}, \bibinfo {author} {\bibfnamefont {J.}~\bibnamefont {Screen}}, \bibinfo {author} {\bibfnamefont {M.}~\bibnamefont {Tjernstr{\"o}m}}, \bibinfo {author} {\bibfnamefont {B.}~\bibnamefont {Newton}}, \bibinfo {author} {\bibfnamefont {X.}~\bibnamefont {Zhang}}, \bibinfo {author} {\bibfnamefont {V.}~\bibnamefont {Popova}}, \bibinfo {author} {\bibfnamefont {C.}~\bibnamefont {Deser}}, \bibinfo {author} {\bibfnamefont {M.}~\bibnamefont {Holland}}, \ and\ \bibinfo {author} {\bibfnamefont {T.}~\bibnamefont {Prowse}},\ }\href@noop {} {\bibfield  {journal} {\bibinfo  {journal} {Journal of Geophysical Research: Biogeosciences}\ }\textbf {\bibinfo {volume} {121}},\ \bibinfo {pages} {586} (\bibinfo {year} {2016})}\BibitemShut {NoStop}%
\bibitem [{\citenamefont {Walsh}\ \emph {et~al.}(2020)\citenamefont {Walsh}, \citenamefont {Ballinger}, \citenamefont {Euskirchen}, \citenamefont {Hanna}, \citenamefont {M{\aa}rd}, \citenamefont {Overland}, \citenamefont {Tangen},\ and\ \citenamefont {Vihma}}]{walsh2020extreme}%
  \BibitemOpen
  \bibfield  {author} {\bibinfo {author} {\bibfnamefont {J.~E.}\ \bibnamefont {Walsh}}, \bibinfo {author} {\bibfnamefont {T.~J.}\ \bibnamefont {Ballinger}}, \bibinfo {author} {\bibfnamefont {E.~S.}\ \bibnamefont {Euskirchen}}, \bibinfo {author} {\bibfnamefont {E.}~\bibnamefont {Hanna}}, \bibinfo {author} {\bibfnamefont {J.}~\bibnamefont {M{\aa}rd}}, \bibinfo {author} {\bibfnamefont {J.~E.}\ \bibnamefont {Overland}}, \bibinfo {author} {\bibfnamefont {H.}~\bibnamefont {Tangen}}, \ and\ \bibinfo {author} {\bibfnamefont {T.}~\bibnamefont {Vihma}},\ }\href@noop {} {\bibfield  {journal} {\bibinfo  {journal} {Earth-Science Reviews}\ }\textbf {\bibinfo {volume} {209}},\ \bibinfo {pages} {103324} (\bibinfo {year} {2020})}\BibitemShut {NoStop}%
\bibitem [{\citenamefont {Albeverio}\ \emph {et~al.}(2006)\citenamefont {Albeverio}, \citenamefont {Jentsch},\ and\ \citenamefont {Kantz}}]{albeverio2006extreme}%
  \BibitemOpen
  \bibfield  {author} {\bibinfo {author} {\bibfnamefont {S.}~\bibnamefont {Albeverio}}, \bibinfo {author} {\bibfnamefont {V.}~\bibnamefont {Jentsch}}, \ and\ \bibinfo {author} {\bibfnamefont {H.}~\bibnamefont {Kantz}},\ }\href@noop {} {\emph {\bibinfo {title} {Extreme events in nature and society}}}\ (\bibinfo  {publisher} {Springer Science \& Business Media},\ \bibinfo {year} {2006})\BibitemShut {NoStop}%
\bibitem [{\citenamefont {P{\"o}rtner}\ \emph {et~al.}(2022)\citenamefont {P{\"o}rtner}, \citenamefont {Roberts}, \citenamefont {Adams}, \citenamefont {Adler}, \citenamefont {Aldunce}, \citenamefont {Ali}, \citenamefont {Begum}, \citenamefont {Betts}, \citenamefont {Kerr}, \citenamefont {Biesbroek} \emph {et~al.}}]{portner2022climate}%
  \BibitemOpen
  \bibfield  {author} {\bibinfo {author} {\bibfnamefont {H.-O.}\ \bibnamefont {P{\"o}rtner}}, \bibinfo {author} {\bibfnamefont {D.~C.}\ \bibnamefont {Roberts}}, \bibinfo {author} {\bibfnamefont {H.}~\bibnamefont {Adams}}, \bibinfo {author} {\bibfnamefont {C.}~\bibnamefont {Adler}}, \bibinfo {author} {\bibfnamefont {P.}~\bibnamefont {Aldunce}}, \bibinfo {author} {\bibfnamefont {E.}~\bibnamefont {Ali}}, \bibinfo {author} {\bibfnamefont {R.~A.}\ \bibnamefont {Begum}}, \bibinfo {author} {\bibfnamefont {R.}~\bibnamefont {Betts}}, \bibinfo {author} {\bibfnamefont {R.~B.}\ \bibnamefont {Kerr}}, \bibinfo {author} {\bibfnamefont {R.}~\bibnamefont {Biesbroek}},  \emph {et~al.},\ }\href@noop {} {\emph {\bibinfo {title} {Climate change 2022: Impacts, adaptation and vulnerability}}}\ (\bibinfo  {publisher} {IPCC Geneva, Switzerland:},\ \bibinfo {year} {2022})\BibitemShut {NoStop}%
\bibitem [{\citenamefont {Ghil}\ \emph {et~al.}(2011)\citenamefont {Ghil}, \citenamefont {Yiou}, \citenamefont {Hallegatte}, \citenamefont {Malamud}, \citenamefont {Naveau}, \citenamefont {Soloviev}, \citenamefont {Friederichs}, \citenamefont {Keilis-Borok}, \citenamefont {Kondrashov}, \citenamefont {Kossobokov} \emph {et~al.}}]{ghil2011extreme}%
  \BibitemOpen
  \bibfield  {author} {\bibinfo {author} {\bibfnamefont {M.}~\bibnamefont {Ghil}}, \bibinfo {author} {\bibfnamefont {P.}~\bibnamefont {Yiou}}, \bibinfo {author} {\bibfnamefont {S.}~\bibnamefont {Hallegatte}}, \bibinfo {author} {\bibfnamefont {B.}~\bibnamefont {Malamud}}, \bibinfo {author} {\bibfnamefont {P.}~\bibnamefont {Naveau}}, \bibinfo {author} {\bibfnamefont {A.}~\bibnamefont {Soloviev}}, \bibinfo {author} {\bibfnamefont {P.}~\bibnamefont {Friederichs}}, \bibinfo {author} {\bibfnamefont {V.}~\bibnamefont {Keilis-Borok}}, \bibinfo {author} {\bibfnamefont {D.}~\bibnamefont {Kondrashov}}, \bibinfo {author} {\bibfnamefont {V.}~\bibnamefont {Kossobokov}},  \emph {et~al.},\ }\href@noop {} {\bibfield  {journal} {\bibinfo  {journal} {Nonlinear Processes in Geophysics}\ }\textbf {\bibinfo {volume} {18}},\ \bibinfo {pages} {295} (\bibinfo {year} {2011})}\BibitemShut {NoStop}%
\bibitem [{\citenamefont {Farazmand}\ and\ \citenamefont {Sapsis}(2019)}]{farazmand2019extreme}%
  \BibitemOpen
  \bibfield  {author} {\bibinfo {author} {\bibfnamefont {M.}~\bibnamefont {Farazmand}}\ and\ \bibinfo {author} {\bibfnamefont {T.~P.}\ \bibnamefont {Sapsis}},\ }\href@noop {} {\bibfield  {journal} {\bibinfo  {journal} {Applied Mechanics Reviews}\ }\textbf {\bibinfo {volume} {71}},\ \bibinfo {pages} {050801} (\bibinfo {year} {2019})}\BibitemShut {NoStop}%
\bibitem [{\citenamefont {Mishra}\ \emph {et~al.}(2020)\citenamefont {Mishra}, \citenamefont {Leo~Kingston}, \citenamefont {Hens}, \citenamefont {Kapitaniak}, \citenamefont {Feudel},\ and\ \citenamefont {Dana}}]{mishra2020routes}%
  \BibitemOpen
  \bibfield  {author} {\bibinfo {author} {\bibfnamefont {A.}~\bibnamefont {Mishra}}, \bibinfo {author} {\bibfnamefont {S.}~\bibnamefont {Leo~Kingston}}, \bibinfo {author} {\bibfnamefont {C.}~\bibnamefont {Hens}}, \bibinfo {author} {\bibfnamefont {T.}~\bibnamefont {Kapitaniak}}, \bibinfo {author} {\bibfnamefont {U.}~\bibnamefont {Feudel}}, \ and\ \bibinfo {author} {\bibfnamefont {S.~K.}\ \bibnamefont {Dana}},\ }\href@noop {} {\bibfield  {journal} {\bibinfo  {journal} {Chaos: An Interdisciplinary Journal of Nonlinear Science}\ }\textbf {\bibinfo {volume} {30}} (\bibinfo {year} {2020})}\BibitemShut {NoStop}%
\bibitem [{\citenamefont {Ray}\ \emph {et~al.}(2020)\citenamefont {Ray}, \citenamefont {Rakshit}, \citenamefont {Basak}, \citenamefont {Dana},\ and\ \citenamefont {Ghosh}}]{ray2020understanding}%
  \BibitemOpen
  \bibfield  {author} {\bibinfo {author} {\bibfnamefont {A.}~\bibnamefont {Ray}}, \bibinfo {author} {\bibfnamefont {S.}~\bibnamefont {Rakshit}}, \bibinfo {author} {\bibfnamefont {G.~K.}\ \bibnamefont {Basak}}, \bibinfo {author} {\bibfnamefont {S.~K.}\ \bibnamefont {Dana}}, \ and\ \bibinfo {author} {\bibfnamefont {D.}~\bibnamefont {Ghosh}},\ }\href@noop {} {\bibfield  {journal} {\bibinfo  {journal} {Physical Review E}\ }\textbf {\bibinfo {volume} {101}},\ \bibinfo {pages} {062210} (\bibinfo {year} {2020})}\BibitemShut {NoStop}%
\bibitem [{\citenamefont {Chowdhury}\ \emph {et~al.}(2021)\citenamefont {Chowdhury}, \citenamefont {Ray}, \citenamefont {Mishra},\ and\ \citenamefont {Ghosh}}]{chowdhury2021extreme}%
  \BibitemOpen
  \bibfield  {author} {\bibinfo {author} {\bibfnamefont {S.~N.}\ \bibnamefont {Chowdhury}}, \bibinfo {author} {\bibfnamefont {A.}~\bibnamefont {Ray}}, \bibinfo {author} {\bibfnamefont {A.}~\bibnamefont {Mishra}}, \ and\ \bibinfo {author} {\bibfnamefont {D.}~\bibnamefont {Ghosh}},\ }\href@noop {} {\bibfield  {journal} {\bibinfo  {journal} {Journal of Physics: Complexity}\ }\textbf {\bibinfo {volume} {2}},\ \bibinfo {pages} {035021} (\bibinfo {year} {2021})}\BibitemShut {NoStop}%
\bibitem [{\citenamefont {Chowdhury}\ \emph {et~al.}(2022)\citenamefont {Chowdhury}, \citenamefont {Ray}, \citenamefont {Dana},\ and\ \citenamefont {Ghosh}}]{chowdhury2022extreme}%
  \BibitemOpen
  \bibfield  {author} {\bibinfo {author} {\bibfnamefont {S.~N.}\ \bibnamefont {Chowdhury}}, \bibinfo {author} {\bibfnamefont {A.}~\bibnamefont {Ray}}, \bibinfo {author} {\bibfnamefont {S.~K.}\ \bibnamefont {Dana}}, \ and\ \bibinfo {author} {\bibfnamefont {D.}~\bibnamefont {Ghosh}},\ }\href@noop {} {\bibfield  {journal} {\bibinfo  {journal} {Physics Reports}\ }\textbf {\bibinfo {volume} {966}},\ \bibinfo {pages} {1} (\bibinfo {year} {2022})}\BibitemShut {NoStop}%
\bibitem [{\citenamefont {Das}\ \emph {et~al.}(2024{\natexlab{a}})\citenamefont {Das}, \citenamefont {Ray}, \citenamefont {Hens}, \citenamefont {Ghosh}, \citenamefont {Hassan}, \citenamefont {Dabrowski}, \citenamefont {Kapitaniak},\ and\ \citenamefont {Dana}}]{das2024complexity}%
  \BibitemOpen
  \bibfield  {author} {\bibinfo {author} {\bibfnamefont {D.}~\bibnamefont {Das}}, \bibinfo {author} {\bibfnamefont {A.}~\bibnamefont {Ray}}, \bibinfo {author} {\bibfnamefont {C.}~\bibnamefont {Hens}}, \bibinfo {author} {\bibfnamefont {D.}~\bibnamefont {Ghosh}}, \bibinfo {author} {\bibfnamefont {M.~K.}\ \bibnamefont {Hassan}}, \bibinfo {author} {\bibfnamefont {A.}~\bibnamefont {Dabrowski}}, \bibinfo {author} {\bibfnamefont {T.}~\bibnamefont {Kapitaniak}}, \ and\ \bibinfo {author} {\bibfnamefont {S.~K.}\ \bibnamefont {Dana}},\ }\href@noop {} {\bibfield  {journal} {\bibinfo  {journal} {Chaos: An Interdisciplinary Journal of Nonlinear Science}\ }\textbf {\bibinfo {volume} {34}} (\bibinfo {year} {2024}{\natexlab{a}})}\BibitemShut {NoStop}%
\bibitem [{\citenamefont {Box}\ \emph {et~al.}(2019)\citenamefont {Box}, \citenamefont {Colgan}, \citenamefont {Christensen}, \citenamefont {Schmidt}, \citenamefont {Lund}, \citenamefont {Parmentier}, \citenamefont {Brown}, \citenamefont {Bhatt}, \citenamefont {Euskirchen}, \citenamefont {Romanovsky} \emph {et~al.}}]{box2019key}%
  \BibitemOpen
  \bibfield  {author} {\bibinfo {author} {\bibfnamefont {J.~E.}\ \bibnamefont {Box}}, \bibinfo {author} {\bibfnamefont {W.~T.}\ \bibnamefont {Colgan}}, \bibinfo {author} {\bibfnamefont {T.~R.}\ \bibnamefont {Christensen}}, \bibinfo {author} {\bibfnamefont {N.~M.}\ \bibnamefont {Schmidt}}, \bibinfo {author} {\bibfnamefont {M.}~\bibnamefont {Lund}}, \bibinfo {author} {\bibfnamefont {F.-J.~W.}\ \bibnamefont {Parmentier}}, \bibinfo {author} {\bibfnamefont {R.}~\bibnamefont {Brown}}, \bibinfo {author} {\bibfnamefont {U.~S.}\ \bibnamefont {Bhatt}}, \bibinfo {author} {\bibfnamefont {E.~S.}\ \bibnamefont {Euskirchen}}, \bibinfo {author} {\bibfnamefont {V.~E.}\ \bibnamefont {Romanovsky}},  \emph {et~al.},\ }\href@noop {} {\bibfield  {journal} {\bibinfo  {journal} {Environmental Research Letters}\ }\textbf {\bibinfo {volume} {14}},\ \bibinfo {pages} {045010} (\bibinfo {year} {2019})}\BibitemShut {NoStop}%
\bibitem [{\citenamefont {Landrum}\ and\ \citenamefont {Holland}(2020)}]{landrum2020extremes}%
  \BibitemOpen
  \bibfield  {author} {\bibinfo {author} {\bibfnamefont {L.}~\bibnamefont {Landrum}}\ and\ \bibinfo {author} {\bibfnamefont {M.~M.}\ \bibnamefont {Holland}},\ }\href@noop {} {\bibfield  {journal} {\bibinfo  {journal} {Nature Climate Change}\ }\textbf {\bibinfo {volume} {10}},\ \bibinfo {pages} {1108} (\bibinfo {year} {2020})}\BibitemShut {NoStop}%
\bibitem [{\citenamefont {Loeng}(1991)}]{loeng1991features}%
  \BibitemOpen
  \bibfield  {author} {\bibinfo {author} {\bibfnamefont {H.}~\bibnamefont {Loeng}},\ }\href@noop {} {\bibfield  {journal} {\bibinfo  {journal} {Polar Research}\ }\textbf {\bibinfo {volume} {10}},\ \bibinfo {pages} {5} (\bibinfo {year} {1991})}\BibitemShut {NoStop}%
\bibitem [{\citenamefont {Loeng}\ \emph {et~al.}(1997)\citenamefont {Loeng}, \citenamefont {Ozhigin},\ and\ \citenamefont {{\AA}dlandsvik}}]{loeng1997water}%
  \BibitemOpen
  \bibfield  {author} {\bibinfo {author} {\bibfnamefont {H.}~\bibnamefont {Loeng}}, \bibinfo {author} {\bibfnamefont {V.}~\bibnamefont {Ozhigin}}, \ and\ \bibinfo {author} {\bibfnamefont {B.}~\bibnamefont {{\AA}dlandsvik}},\ }\href@noop {} {\bibfield  {journal} {\bibinfo  {journal} {ICES Journal of Marine Science}\ }\textbf {\bibinfo {volume} {54}},\ \bibinfo {pages} {310} (\bibinfo {year} {1997})}\BibitemShut {NoStop}%
\bibitem [{\citenamefont {Carroll}\ \emph {et~al.}(2011)\citenamefont {Carroll}, \citenamefont {Ambrose~Jr}, \citenamefont {Levin}, \citenamefont {Henkes}, \citenamefont {Hop}, \citenamefont {Renaud} \emph {et~al.}}]{carroll2011pan}%
  \BibitemOpen
  \bibfield  {author} {\bibinfo {author} {\bibfnamefont {M.~L.}\ \bibnamefont {Carroll}}, \bibinfo {author} {\bibfnamefont {W.~G.}\ \bibnamefont {Ambrose~Jr}}, \bibinfo {author} {\bibfnamefont {B.~S.}\ \bibnamefont {Levin}}, \bibinfo {author} {\bibfnamefont {G.~A.}\ \bibnamefont {Henkes}}, \bibinfo {author} {\bibfnamefont {H.}~\bibnamefont {Hop}}, \bibinfo {author} {\bibfnamefont {P.~E.}\ \bibnamefont {Renaud}},  \emph {et~al.},\ }\href@noop {} {\bibfield  {journal} {\bibinfo  {journal} {Journal of Marine Systems}\ }\textbf {\bibinfo {volume} {88}},\ \bibinfo {pages} {239} (\bibinfo {year} {2011})}\BibitemShut {NoStop}%
\bibitem [{\citenamefont {{\L}upikasza}\ \emph {et~al.}(2021)\citenamefont {{\L}upikasza}, \citenamefont {Nied{\'z}wied{\'z}}, \citenamefont {Przybylak},\ and\ \citenamefont {Nordli}}]{lupikasza2021importance}%
  \BibitemOpen
  \bibfield  {author} {\bibinfo {author} {\bibfnamefont {E.~B.}\ \bibnamefont {{\L}upikasza}}, \bibinfo {author} {\bibfnamefont {T.}~\bibnamefont {Nied{\'z}wied{\'z}}}, \bibinfo {author} {\bibfnamefont {R.}~\bibnamefont {Przybylak}}, \ and\ \bibinfo {author} {\bibfnamefont {{\O}.}~\bibnamefont {Nordli}},\ }\href@noop {} {\bibfield  {journal} {\bibinfo  {journal} {International Journal of Climatology}\ }\textbf {\bibinfo {volume} {41}},\ \bibinfo {pages} {3481} (\bibinfo {year} {2021})}\BibitemShut {NoStop}%
\bibitem [{\citenamefont {Maturilli}\ and\ \citenamefont {Kayser}(2017)}]{maturilli2017arctic}%
  \BibitemOpen
  \bibfield  {author} {\bibinfo {author} {\bibfnamefont {M.}~\bibnamefont {Maturilli}}\ and\ \bibinfo {author} {\bibfnamefont {M.}~\bibnamefont {Kayser}},\ }\href@noop {} {\bibfield  {journal} {\bibinfo  {journal} {Theoretical and Applied Climatology}\ }\textbf {\bibinfo {volume} {130}},\ \bibinfo {pages} {1} (\bibinfo {year} {2017})}\BibitemShut {NoStop}%
\bibitem [{\citenamefont {Platt}\ \emph {et~al.}(2021)\citenamefont {Platt}, \citenamefont {Hov}, \citenamefont {Berg}, \citenamefont {Breivik}, \citenamefont {Eckhardt}, \citenamefont {Eleftheriadis}, \citenamefont {Evangeliou}, \citenamefont {Fiebig}, \citenamefont {Fisher}, \citenamefont {Hansen} \emph {et~al.}}]{platt2021atmospheric}%
  \BibitemOpen
  \bibfield  {author} {\bibinfo {author} {\bibfnamefont {S.~M.}\ \bibnamefont {Platt}}, \bibinfo {author} {\bibfnamefont {{\O}.}~\bibnamefont {Hov}}, \bibinfo {author} {\bibfnamefont {T.}~\bibnamefont {Berg}}, \bibinfo {author} {\bibfnamefont {K.}~\bibnamefont {Breivik}}, \bibinfo {author} {\bibfnamefont {S.}~\bibnamefont {Eckhardt}}, \bibinfo {author} {\bibfnamefont {K.}~\bibnamefont {Eleftheriadis}}, \bibinfo {author} {\bibfnamefont {N.}~\bibnamefont {Evangeliou}}, \bibinfo {author} {\bibfnamefont {M.}~\bibnamefont {Fiebig}}, \bibinfo {author} {\bibfnamefont {R.}~\bibnamefont {Fisher}}, \bibinfo {author} {\bibfnamefont {G.}~\bibnamefont {Hansen}},  \emph {et~al.},\ }\href@noop {} {\bibfield  {journal} {\bibinfo  {journal} {Atmospheric Chemistry and Physics Discussions}\ }\textbf {\bibinfo {volume} {2021}},\ \bibinfo {pages} {1} (\bibinfo {year} {2021})}\BibitemShut {NoStop}%
\bibitem [{\citenamefont {Owczarek}\ \emph {et~al.}(2020)\citenamefont {Owczarek}, \citenamefont {Opa{\l}a-Owczarek},\ and\ \citenamefont {Miga{\l}a}}]{owczarek2020post}%
  \BibitemOpen
  \bibfield  {author} {\bibinfo {author} {\bibfnamefont {P.}~\bibnamefont {Owczarek}}, \bibinfo {author} {\bibfnamefont {M.}~\bibnamefont {Opa{\l}a-Owczarek}}, \ and\ \bibinfo {author} {\bibfnamefont {K.}~\bibnamefont {Miga{\l}a}},\ }\href@noop {} {\bibfield  {journal} {\bibinfo  {journal} {Environmental Research Letters}\ }\textbf {\bibinfo {volume} {16}},\ \bibinfo {pages} {014031} (\bibinfo {year} {2020})}\BibitemShut {NoStop}%
\bibitem [{\citenamefont {Ren}\ \emph {et~al.}(2021)\citenamefont {Ren}, \citenamefont {Wang}, \citenamefont {Lu}, \citenamefont {Wu}, \citenamefont {Lu}, \citenamefont {Chen},\ and\ \citenamefont {Ma}}]{rs13193845}%
  \BibitemOpen
  \bibfield  {author} {\bibinfo {author} {\bibfnamefont {G.}~\bibnamefont {Ren}}, \bibinfo {author} {\bibfnamefont {J.}~\bibnamefont {Wang}}, \bibinfo {author} {\bibfnamefont {Y.}~\bibnamefont {Lu}}, \bibinfo {author} {\bibfnamefont {P.}~\bibnamefont {Wu}}, \bibinfo {author} {\bibfnamefont {X.}~\bibnamefont {Lu}}, \bibinfo {author} {\bibfnamefont {C.}~\bibnamefont {Chen}}, \ and\ \bibinfo {author} {\bibfnamefont {Y.}~\bibnamefont {Ma}},\ }\href {\doibase 10.3390/rs13193845} {\bibfield  {journal} {\bibinfo  {journal} {Remote Sensing}\ }\textbf {\bibinfo {volume} {13}} (\bibinfo {year} {2021}),\ 10.3390/rs13193845}\BibitemShut {NoStop}%
\bibitem [{\citenamefont {Putkonen}(1998)}]{Putkonen12011998}%
  \BibitemOpen
  \bibfield  {author} {\bibinfo {author} {\bibfnamefont {J.}~\bibnamefont {Putkonen}},\ }\href {\doibase 10.3402/polar.v17i2.6617} {\bibfield  {journal} {\bibinfo  {journal} {Polar Research}\ }\textbf {\bibinfo {volume} {17}},\ \bibinfo {pages} {165} (\bibinfo {year} {1998})}\BibitemShut {NoStop}%
\bibitem [{\citenamefont {Das}\ \emph {et~al.}(2025)\citenamefont {Das}, \citenamefont {Athulya}, \citenamefont {Chakraborty}, \citenamefont {Ray}, \citenamefont {Hens}, \citenamefont {Dana}, \citenamefont {Ghosh},\ and\ \citenamefont {Murukesh}}]{das2025pattern}%
  \BibitemOpen
  \bibfield  {author} {\bibinfo {author} {\bibfnamefont {D.}~\bibnamefont {Das}}, \bibinfo {author} {\bibfnamefont {R.}~\bibnamefont {Athulya}}, \bibinfo {author} {\bibfnamefont {T.}~\bibnamefont {Chakraborty}}, \bibinfo {author} {\bibfnamefont {A.}~\bibnamefont {Ray}}, \bibinfo {author} {\bibfnamefont {C.}~\bibnamefont {Hens}}, \bibinfo {author} {\bibfnamefont {S.~K.}\ \bibnamefont {Dana}}, \bibinfo {author} {\bibfnamefont {D.}~\bibnamefont {Ghosh}}, \ and\ \bibinfo {author} {\bibfnamefont {N.}~\bibnamefont {Murukesh}},\ }\href@noop {} {\bibfield  {journal} {\bibinfo  {journal} {Scientific Reports}\ }\textbf {\bibinfo {volume} {15}},\ \bibinfo {pages} {8754} (\bibinfo {year} {2025})}\BibitemShut {NoStop}%
\bibitem [{\citenamefont {Czernecki}\ \emph {et~al.}(2015)\citenamefont {Czernecki}, \citenamefont {Taszarek}, \citenamefont {Kolendowicz},\ and\ \citenamefont {Szyga-Pluta}}]{czernecki2015atmospheric}%
  \BibitemOpen
  \bibfield  {author} {\bibinfo {author} {\bibfnamefont {B.}~\bibnamefont {Czernecki}}, \bibinfo {author} {\bibfnamefont {M.}~\bibnamefont {Taszarek}}, \bibinfo {author} {\bibfnamefont {L.}~\bibnamefont {Kolendowicz}}, \ and\ \bibinfo {author} {\bibfnamefont {K.}~\bibnamefont {Szyga-Pluta}},\ }\href@noop {} {\bibfield  {journal} {\bibinfo  {journal} {Atmospheric Research}\ }\textbf {\bibinfo {volume} {154}},\ \bibinfo {pages} {60} (\bibinfo {year} {2015})}\BibitemShut {NoStop}%
\bibitem [{\citenamefont {Nordeng}\ and\ \citenamefont {Rasmussen}(1992)}]{nordeng1992most}%
  \BibitemOpen
  \bibfield  {author} {\bibinfo {author} {\bibfnamefont {T.~E.}\ \bibnamefont {Nordeng}}\ and\ \bibinfo {author} {\bibfnamefont {E.~A.}\ \bibnamefont {Rasmussen}},\ }\href@noop {} {\bibfield  {journal} {\bibinfo  {journal} {Tellus A}\ }\textbf {\bibinfo {volume} {44}},\ \bibinfo {pages} {81} (\bibinfo {year} {1992})}\BibitemShut {NoStop}%
\bibitem [{\citenamefont {Wendisch}\ \emph {et~al.}(2019)\citenamefont {Wendisch}, \citenamefont {Macke}, \citenamefont {Ehrlich}, \citenamefont {L{\"u}pkes}, \citenamefont {Mech}, \citenamefont {Chechin}, \citenamefont {Dethloff}, \citenamefont {Velasco}, \citenamefont {Bozem}, \citenamefont {Br{\"u}ckner} \emph {et~al.}}]{wendisch2019arctic}%
  \BibitemOpen
  \bibfield  {author} {\bibinfo {author} {\bibfnamefont {M.}~\bibnamefont {Wendisch}}, \bibinfo {author} {\bibfnamefont {A.}~\bibnamefont {Macke}}, \bibinfo {author} {\bibfnamefont {A.}~\bibnamefont {Ehrlich}}, \bibinfo {author} {\bibfnamefont {C.}~\bibnamefont {L{\"u}pkes}}, \bibinfo {author} {\bibfnamefont {M.}~\bibnamefont {Mech}}, \bibinfo {author} {\bibfnamefont {D.}~\bibnamefont {Chechin}}, \bibinfo {author} {\bibfnamefont {K.}~\bibnamefont {Dethloff}}, \bibinfo {author} {\bibfnamefont {C.~B.}\ \bibnamefont {Velasco}}, \bibinfo {author} {\bibfnamefont {H.}~\bibnamefont {Bozem}}, \bibinfo {author} {\bibfnamefont {M.}~\bibnamefont {Br{\"u}ckner}},  \emph {et~al.},\ }\href@noop {} {\bibfield  {journal} {\bibinfo  {journal} {Bulletin of the American Meteorological Society}\ }\textbf {\bibinfo {volume} {100}},\ \bibinfo {pages} {841} (\bibinfo {year} {2019})}\BibitemShut {NoStop}%
\bibitem [{\citenamefont {Pernov}\ \emph {et~al.}(2022)\citenamefont {Pernov}, \citenamefont {Beddows}, \citenamefont {Thomas},\ and\ \citenamefont {et~al.}}]{Pernov2022}%
  \BibitemOpen
  \bibfield  {author} {\bibinfo {author} {\bibfnamefont {J.~B.}\ \bibnamefont {Pernov}}, \bibinfo {author} {\bibfnamefont {D.}~\bibnamefont {Beddows}}, \bibinfo {author} {\bibfnamefont {D.~C.}\ \bibnamefont {Thomas}}, \ and\ \bibinfo {author} {\bibnamefont {et~al.}},\ }\href {\doibase 10.1038/s41612-022-00286-y} {\bibfield  {journal} {\bibinfo  {journal} {npj Climate and Atmospheric Science}\ }\textbf {\bibinfo {volume} {5}},\ \bibinfo {pages} {62} (\bibinfo {year} {2022})}\BibitemShut {NoStop}%
\bibitem [{\citenamefont {Acharya}\ \emph {et~al.}(2021)\citenamefont {Acharya}, \citenamefont {Chatterjee}, \citenamefont {Subeesh}, \citenamefont {Radhakrishnan},\ and\ \citenamefont {Murukesh}}]{asutosh2021}%
  \BibitemOpen
  \bibfield  {author} {\bibinfo {author} {\bibfnamefont {A.}~\bibnamefont {Acharya}}, \bibinfo {author} {\bibfnamefont {S.}~\bibnamefont {Chatterjee}}, \bibinfo {author} {\bibfnamefont {M.~P.}\ \bibnamefont {Subeesh}}, \bibinfo {author} {\bibfnamefont {A.}~\bibnamefont {Radhakrishnan}}, \ and\ \bibinfo {author} {\bibfnamefont {N.}~\bibnamefont {Murukesh}},\ }\href {\doibase 10.3390/rs13142808} {\bibfield  {journal} {\bibinfo  {journal} {Remote Sensing}\ }\textbf {\bibinfo {volume} {13}},\ \bibinfo {pages} {2808} (\bibinfo {year} {2021})}\BibitemShut {NoStop}%
\bibitem [{\citenamefont {Athulya}\ \emph {et~al.}(2023)\citenamefont {Athulya}, \citenamefont {Nuncio}, \citenamefont {Chatterjee},\ and\ \citenamefont {Vidya}}]{ATHULYA2023106989}%
  \BibitemOpen
  \bibfield  {author} {\bibinfo {author} {\bibfnamefont {R.}~\bibnamefont {Athulya}}, \bibinfo {author} {\bibfnamefont {M.}~\bibnamefont {Nuncio}}, \bibinfo {author} {\bibfnamefont {S.}~\bibnamefont {Chatterjee}}, \ and\ \bibinfo {author} {\bibfnamefont {P.}~\bibnamefont {Vidya}},\ }\href {\doibase https://doi.org/10.1016/j.atmosres.2023.106989} {\bibfield  {journal} {\bibinfo  {journal} {Atmospheric Research}\ }\textbf {\bibinfo {volume} {295}},\ \bibinfo {pages} {106989} (\bibinfo {year} {2023})}\BibitemShut {NoStop}%
\bibitem [{\citenamefont {Torrence}\ and\ \citenamefont {Compo}(1998)}]{torrence1998practical}%
  \BibitemOpen
  \bibfield  {author} {\bibinfo {author} {\bibfnamefont {C.}~\bibnamefont {Torrence}}\ and\ \bibinfo {author} {\bibfnamefont {G.~P.}\ \bibnamefont {Compo}},\ }\href@noop {} {\bibfield  {journal} {\bibinfo  {journal} {Bulletin of the American Meteorological society}\ }\textbf {\bibinfo {volume} {79}},\ \bibinfo {pages} {61} (\bibinfo {year} {1998})}\BibitemShut {NoStop}%
\bibitem [{\citenamefont {Grinsted}\ \emph {et~al.}(2004)\citenamefont {Grinsted}, \citenamefont {Moore},\ and\ \citenamefont {Jevrejeva}}]{grinsted2004application}%
  \BibitemOpen
  \bibfield  {author} {\bibinfo {author} {\bibfnamefont {A.}~\bibnamefont {Grinsted}}, \bibinfo {author} {\bibfnamefont {J.~C.}\ \bibnamefont {Moore}}, \ and\ \bibinfo {author} {\bibfnamefont {S.}~\bibnamefont {Jevrejeva}},\ }\href@noop {} {\bibfield  {journal} {\bibinfo  {journal} {Nonlinear Processes in Geophysics}\ }\textbf {\bibinfo {volume} {11}},\ \bibinfo {pages} {561} (\bibinfo {year} {2004})}\BibitemShut {NoStop}%
\bibitem [{\citenamefont {Stroeve}\ \emph {et~al.}(2012)\citenamefont {Stroeve}, \citenamefont {Serreze}, \citenamefont {Holland}, \citenamefont {Kay}, \citenamefont {Malanik},\ and\ \citenamefont {Barrett}}]{stroeve2012arctic}%
  \BibitemOpen
  \bibfield  {author} {\bibinfo {author} {\bibfnamefont {J.~C.}\ \bibnamefont {Stroeve}}, \bibinfo {author} {\bibfnamefont {M.~C.}\ \bibnamefont {Serreze}}, \bibinfo {author} {\bibfnamefont {M.~M.}\ \bibnamefont {Holland}}, \bibinfo {author} {\bibfnamefont {J.~E.}\ \bibnamefont {Kay}}, \bibinfo {author} {\bibfnamefont {J.}~\bibnamefont {Malanik}}, \ and\ \bibinfo {author} {\bibfnamefont {A.~P.}\ \bibnamefont {Barrett}},\ }\href@noop {} {\bibfield  {journal} {\bibinfo  {journal} {Climatic Change}\ }\textbf {\bibinfo {volume} {110}},\ \bibinfo {pages} {1005} (\bibinfo {year} {2012})}\BibitemShut {NoStop}%
\bibitem [{\citenamefont {Mart{\'\i}nez-S{\'a}nchez}\ \emph {et~al.}(2024)\citenamefont {Mart{\'\i}nez-S{\'a}nchez}, \citenamefont {Arranz},\ and\ \citenamefont {Lozano-Dur{\'a}n}}]{martinez2024decomposing}%
  \BibitemOpen
  \bibfield  {author} {\bibinfo {author} {\bibfnamefont {{\'A}.}~\bibnamefont {Mart{\'\i}nez-S{\'a}nchez}}, \bibinfo {author} {\bibfnamefont {G.}~\bibnamefont {Arranz}}, \ and\ \bibinfo {author} {\bibfnamefont {A.}~\bibnamefont {Lozano-Dur{\'a}n}},\ }\href@noop {} {\bibfield  {journal} {\bibinfo  {journal} {Nature Communications}\ }\textbf {\bibinfo {volume} {15}},\ \bibinfo {pages} {9296} (\bibinfo {year} {2024})}\BibitemShut {NoStop}%
\bibitem [{\citenamefont {Benziane}\ and\ \citenamefont {MB}(2024)}]{benziane2024survey}%
  \BibitemOpen
  \bibfield  {author} {\bibinfo {author} {\bibfnamefont {S.}~\bibnamefont {Benziane}}\ and\ \bibinfo {author} {\bibfnamefont {U.}~\bibnamefont {MB}},\ }\href@noop {} {\bibfield  {journal} {\bibinfo  {journal} {WSEAS Transactions on Systems}\ }\textbf {\bibinfo {volume} {23}},\ \bibinfo {pages} {47} (\bibinfo {year} {2024})}\BibitemShut {NoStop}%
\bibitem [{\citenamefont {Lai}\ and\ \citenamefont {Dzombak}(2020)}]{lai2020use}%
  \BibitemOpen
  \bibfield  {author} {\bibinfo {author} {\bibfnamefont {Y.}~\bibnamefont {Lai}}\ and\ \bibinfo {author} {\bibfnamefont {D.~A.}\ \bibnamefont {Dzombak}},\ }\href@noop {} {\bibfield  {journal} {\bibinfo  {journal} {Weather and Forecasting}\ }\textbf {\bibinfo {volume} {35}},\ \bibinfo {pages} {959} (\bibinfo {year} {2020})}\BibitemShut {NoStop}%
\bibitem [{\citenamefont {Bari}\ \emph {et~al.}(2015)\citenamefont {Bari}, \citenamefont {Rahman}, \citenamefont {Hussain},\ and\ \citenamefont {Ray}}]{bari2015forecasting}%
  \BibitemOpen
  \bibfield  {author} {\bibinfo {author} {\bibfnamefont {S.~H.}\ \bibnamefont {Bari}}, \bibinfo {author} {\bibfnamefont {M.}~\bibnamefont {Rahman}}, \bibinfo {author} {\bibfnamefont {M.}~\bibnamefont {Hussain}}, \ and\ \bibinfo {author} {\bibfnamefont {S.}~\bibnamefont {Ray}},\ }\href@noop {} {\bibfield  {journal} {\bibinfo  {journal} {Civil and Environmental Research}\ }\textbf {\bibinfo {volume} {7}},\ \bibinfo {pages} {69} (\bibinfo {year} {2015})}\BibitemShut {NoStop}%
\bibitem [{\citenamefont {Herman}\ and\ \citenamefont {Schumacher}(2018)}]{herman2018money}%
  \BibitemOpen
  \bibfield  {author} {\bibinfo {author} {\bibfnamefont {G.~R.}\ \bibnamefont {Herman}}\ and\ \bibinfo {author} {\bibfnamefont {R.~S.}\ \bibnamefont {Schumacher}},\ }\href@noop {} {\bibfield  {journal} {\bibinfo  {journal} {Monthly Weather Review}\ }\textbf {\bibinfo {volume} {146}},\ \bibinfo {pages} {1571} (\bibinfo {year} {2018})}\BibitemShut {NoStop}%
\bibitem [{\citenamefont {Hall}\ \emph {et~al.}(1999)\citenamefont {Hall}, \citenamefont {Brooks},\ and\ \citenamefont {Doswell~Iii}}]{hall1999precipitation}%
  \BibitemOpen
  \bibfield  {author} {\bibinfo {author} {\bibfnamefont {T.}~\bibnamefont {Hall}}, \bibinfo {author} {\bibfnamefont {H.~E.}\ \bibnamefont {Brooks}}, \ and\ \bibinfo {author} {\bibfnamefont {C.~A.}\ \bibnamefont {Doswell~Iii}},\ }\href@noop {} {\bibfield  {journal} {\bibinfo  {journal} {Weather and Forecasting}\ }\textbf {\bibinfo {volume} {14}},\ \bibinfo {pages} {338} (\bibinfo {year} {1999})}\BibitemShut {NoStop}%
\bibitem [{\citenamefont {Shi}\ \emph {et~al.}(2015)\citenamefont {Shi}, \citenamefont {Chen}, \citenamefont {Wang}, \citenamefont {Yeung}, \citenamefont {Wong},\ and\ \citenamefont {Woo}}]{shi2015convolutional}%
  \BibitemOpen
  \bibfield  {author} {\bibinfo {author} {\bibfnamefont {X.}~\bibnamefont {Shi}}, \bibinfo {author} {\bibfnamefont {Z.}~\bibnamefont {Chen}}, \bibinfo {author} {\bibfnamefont {H.}~\bibnamefont {Wang}}, \bibinfo {author} {\bibfnamefont {D.-Y.}\ \bibnamefont {Yeung}}, \bibinfo {author} {\bibfnamefont {W.-K.}\ \bibnamefont {Wong}}, \ and\ \bibinfo {author} {\bibfnamefont {W.-c.}\ \bibnamefont {Woo}},\ }\href@noop {} {\bibfield  {journal} {\bibinfo  {journal} {Advances in Neural Information Processing Systems}\ }\textbf {\bibinfo {volume} {28}} (\bibinfo {year} {2015})}\BibitemShut {NoStop}%
\bibitem [{\citenamefont {Das}\ \emph {et~al.}(2024{\natexlab{b}})\citenamefont {Das}, \citenamefont {Posch}, \citenamefont {Barber}, \citenamefont {Hicks}, \citenamefont {Duffy}, \citenamefont {Vandal}, \citenamefont {Singh}, \citenamefont {Werkhoven},\ and\ \citenamefont {Ganguly}}]{das2024hybrid}%
  \BibitemOpen
  \bibfield  {author} {\bibinfo {author} {\bibfnamefont {P.}~\bibnamefont {Das}}, \bibinfo {author} {\bibfnamefont {A.}~\bibnamefont {Posch}}, \bibinfo {author} {\bibfnamefont {N.}~\bibnamefont {Barber}}, \bibinfo {author} {\bibfnamefont {M.}~\bibnamefont {Hicks}}, \bibinfo {author} {\bibfnamefont {K.}~\bibnamefont {Duffy}}, \bibinfo {author} {\bibfnamefont {T.}~\bibnamefont {Vandal}}, \bibinfo {author} {\bibfnamefont {D.}~\bibnamefont {Singh}}, \bibinfo {author} {\bibfnamefont {K.~v.}\ \bibnamefont {Werkhoven}}, \ and\ \bibinfo {author} {\bibfnamefont {A.~R.}\ \bibnamefont {Ganguly}},\ }\href@noop {} {\bibfield  {journal} {\bibinfo  {journal} {npj Climate and Atmospheric Science}\ }\textbf {\bibinfo {volume} {7}},\ \bibinfo {pages} {282} (\bibinfo {year} {2024}{\natexlab{b}})}\BibitemShut {NoStop}%
\bibitem [{\citenamefont {Xu}\ \emph {et~al.}(2024)\citenamefont {Xu}, \citenamefont {Qin}, \citenamefont {Sun}, \citenamefont {Liao},\ and\ \citenamefont {Zheng}}]{xu2024pfformer}%
  \BibitemOpen
  \bibfield  {author} {\bibinfo {author} {\bibfnamefont {L.}~\bibnamefont {Xu}}, \bibinfo {author} {\bibfnamefont {J.}~\bibnamefont {Qin}}, \bibinfo {author} {\bibfnamefont {D.}~\bibnamefont {Sun}}, \bibinfo {author} {\bibfnamefont {Y.}~\bibnamefont {Liao}}, \ and\ \bibinfo {author} {\bibfnamefont {J.}~\bibnamefont {Zheng}},\ }\href@noop {} {\bibfield  {journal} {\bibinfo  {journal} {IEEE Access}\ } (\bibinfo {year} {2024})}\BibitemShut {NoStop}%
\bibitem [{\citenamefont {Worslex}\ \emph {et~al.}(2001)\citenamefont {Worslex}, \citenamefont {Agdestein}, \citenamefont {Gjelberg}, \citenamefont {Kirkemo}, \citenamefont {M{\o}rk}, \citenamefont {Nilsson}, \citenamefont {Olaussen}, \citenamefont {Steel},\ and\ \citenamefont {Stemmerik}}]{worslex2001geological}%
  \BibitemOpen
  \bibfield  {author} {\bibinfo {author} {\bibfnamefont {D.}~\bibnamefont {Worslex}}, \bibinfo {author} {\bibfnamefont {T.}~\bibnamefont {Agdestein}}, \bibinfo {author} {\bibfnamefont {J.~G.}\ \bibnamefont {Gjelberg}}, \bibinfo {author} {\bibfnamefont {K.}~\bibnamefont {Kirkemo}}, \bibinfo {author} {\bibfnamefont {A.}~\bibnamefont {M{\o}rk}}, \bibinfo {author} {\bibfnamefont {I.}~\bibnamefont {Nilsson}}, \bibinfo {author} {\bibfnamefont {S.}~\bibnamefont {Olaussen}}, \bibinfo {author} {\bibfnamefont {R.~J.}\ \bibnamefont {Steel}}, \ and\ \bibinfo {author} {\bibfnamefont {L.}~\bibnamefont {Stemmerik}},\ }\href@noop {} {\bibfield  {journal} {\bibinfo  {journal} {Norwegian Journal of Geology/Norsk Geologisk Forening}\ }\textbf {\bibinfo {volume} {81}} (\bibinfo {year} {2001})}\BibitemShut {NoStop}%
\bibitem [{\citenamefont {Svendsen}\ \emph {et~al.}(2002)\citenamefont {Svendsen}, \citenamefont {Beszczynska-Møller}, \citenamefont {Hagen}, \citenamefont {Lefauconnier}, \citenamefont {Tverberg}, \citenamefont {Gerland}, \citenamefont {Ørbæk}, \citenamefont {Bischof}, \citenamefont {Papucci}, \citenamefont {Zajaczkowski}, \citenamefont {Azzolini}, \citenamefont {Bruland},\ and\ \citenamefont {Wiencke}}]{Svendsen06012002}%
  \BibitemOpen
  \bibfield  {author} {\bibinfo {author} {\bibfnamefont {H.}~\bibnamefont {Svendsen}}, \bibinfo {author} {\bibfnamefont {A.}~\bibnamefont {Beszczynska-Møller}}, \bibinfo {author} {\bibfnamefont {J.~O.}\ \bibnamefont {Hagen}}, \bibinfo {author} {\bibfnamefont {B.}~\bibnamefont {Lefauconnier}}, \bibinfo {author} {\bibfnamefont {V.}~\bibnamefont {Tverberg}}, \bibinfo {author} {\bibfnamefont {S.}~\bibnamefont {Gerland}}, \bibinfo {author} {\bibfnamefont {J.~B.}\ \bibnamefont {Ørbæk}}, \bibinfo {author} {\bibfnamefont {K.}~\bibnamefont {Bischof}}, \bibinfo {author} {\bibfnamefont {C.}~\bibnamefont {Papucci}}, \bibinfo {author} {\bibfnamefont {M.}~\bibnamefont {Zajaczkowski}}, \bibinfo {author} {\bibfnamefont {R.}~\bibnamefont {Azzolini}}, \bibinfo {author} {\bibfnamefont {O.}~\bibnamefont {Bruland}}, \ and\ \bibinfo {author} {\bibfnamefont {C.}~\bibnamefont {Wiencke}},\ }\href {\doibase 10.3402/polar.v21i1.6479} {\bibfield  {journal} {\bibinfo  {journal} {Polar Research}\ }\textbf {\bibinfo {volume} {21}},\
  \bibinfo {pages} {133} (\bibinfo {year} {2002})}\BibitemShut {NoStop}%
\bibitem [{\citenamefont {Richards}\ and\ \citenamefont {Arkin}(1981)}]{richards1981relationship}%
  \BibitemOpen
  \bibfield  {author} {\bibinfo {author} {\bibfnamefont {F.}~\bibnamefont {Richards}}\ and\ \bibinfo {author} {\bibfnamefont {P.}~\bibnamefont {Arkin}},\ }\href@noop {} {\bibfield  {journal} {\bibinfo  {journal} {Monthly Weather Review}\ }\textbf {\bibinfo {volume} {109}},\ \bibinfo {pages} {1081} (\bibinfo {year} {1981})}\BibitemShut {NoStop}%
\bibitem [{\citenamefont {Wang}\ \emph {et~al.}(2018)\citenamefont {Wang}, \citenamefont {Sun},\ and\ \citenamefont {Liu}}]{wang2018dependence}%
  \BibitemOpen
  \bibfield  {author} {\bibinfo {author} {\bibfnamefont {H.}~\bibnamefont {Wang}}, \bibinfo {author} {\bibfnamefont {F.}~\bibnamefont {Sun}}, \ and\ \bibinfo {author} {\bibfnamefont {W.}~\bibnamefont {Liu}},\ }\href@noop {} {\bibfield  {journal} {\bibinfo  {journal} {Journal of Climate}\ }\textbf {\bibinfo {volume} {31}},\ \bibinfo {pages} {8931} (\bibinfo {year} {2018})}\BibitemShut {NoStop}%
\bibitem [{\citenamefont {Yu}\ \emph {et~al.}(2018)\citenamefont {Yu}, \citenamefont {Miller}, \citenamefont {Montalto},\ and\ \citenamefont {Lall}}]{yu2018bridge}%
  \BibitemOpen
  \bibfield  {author} {\bibinfo {author} {\bibfnamefont {Z.}~\bibnamefont {Yu}}, \bibinfo {author} {\bibfnamefont {S.}~\bibnamefont {Miller}}, \bibinfo {author} {\bibfnamefont {F.}~\bibnamefont {Montalto}}, \ and\ \bibinfo {author} {\bibfnamefont {U.}~\bibnamefont {Lall}},\ }\href@noop {} {\bibfield  {journal} {\bibinfo  {journal} {Journal of Hydrology}\ }\textbf {\bibinfo {volume} {562}},\ \bibinfo {pages} {346} (\bibinfo {year} {2018})}\BibitemShut {NoStop}%
\bibitem [{\citenamefont {Panja}\ \emph {et~al.}(2023)\citenamefont {Panja}, \citenamefont {Chakraborty}, \citenamefont {Nadim}, \citenamefont {Ghosh}, \citenamefont {Kumar},\ and\ \citenamefont {Liu}}]{panja2023ensemble}%
  \BibitemOpen
  \bibfield  {author} {\bibinfo {author} {\bibfnamefont {M.}~\bibnamefont {Panja}}, \bibinfo {author} {\bibfnamefont {T.}~\bibnamefont {Chakraborty}}, \bibinfo {author} {\bibfnamefont {S.~S.}\ \bibnamefont {Nadim}}, \bibinfo {author} {\bibfnamefont {I.}~\bibnamefont {Ghosh}}, \bibinfo {author} {\bibfnamefont {U.}~\bibnamefont {Kumar}}, \ and\ \bibinfo {author} {\bibfnamefont {N.}~\bibnamefont {Liu}},\ }\href@noop {} {\bibfield  {journal} {\bibinfo  {journal} {Chaos, Solitons \& Fractals}\ }\textbf {\bibinfo {volume} {167}},\ \bibinfo {pages} {113124} (\bibinfo {year} {2023})}\BibitemShut {NoStop}%
\bibitem [{\citenamefont {Zeng}\ \emph {et~al.}(2023)\citenamefont {Zeng}, \citenamefont {Chen}, \citenamefont {Zhang},\ and\ \citenamefont {Xu}}]{zeng2023transformers}%
  \BibitemOpen
  \bibfield  {author} {\bibinfo {author} {\bibfnamefont {A.}~\bibnamefont {Zeng}}, \bibinfo {author} {\bibfnamefont {M.}~\bibnamefont {Chen}}, \bibinfo {author} {\bibfnamefont {L.}~\bibnamefont {Zhang}}, \ and\ \bibinfo {author} {\bibfnamefont {Q.}~\bibnamefont {Xu}},\ }in\ \href@noop {} {\emph {\bibinfo {booktitle} {Proceedings of the AAAI conference on artificial intelligence}}},\ \bibinfo {series and number} {\bibinfo {number} {9}}\ (\bibinfo {year} {2023})\ pp.\ \bibinfo {pages} {11121--11128}\BibitemShut {NoStop}%
\bibitem [{\citenamefont {Breiman}(2001)}]{breiman2001random}%
  \BibitemOpen
  \bibfield  {author} {\bibinfo {author} {\bibfnamefont {L.}~\bibnamefont {Breiman}},\ }\href@noop {} {\bibfield  {journal} {\bibinfo  {journal} {Machine Learning}\ }\textbf {\bibinfo {volume} {45}},\ \bibinfo {pages} {5} (\bibinfo {year} {2001})}\BibitemShut {NoStop}%
\bibitem [{\citenamefont {Goehry}\ \emph {et~al.}(2023)\citenamefont {Goehry}, \citenamefont {Yan}, \citenamefont {Goude}, \citenamefont {Massart},\ and\ \citenamefont {Poggi}}]{goehry2023random}%
  \BibitemOpen
  \bibfield  {author} {\bibinfo {author} {\bibfnamefont {B.}~\bibnamefont {Goehry}}, \bibinfo {author} {\bibfnamefont {H.}~\bibnamefont {Yan}}, \bibinfo {author} {\bibfnamefont {Y.}~\bibnamefont {Goude}}, \bibinfo {author} {\bibfnamefont {P.}~\bibnamefont {Massart}}, \ and\ \bibinfo {author} {\bibfnamefont {J.-M.}\ \bibnamefont {Poggi}},\ }\href@noop {} {\bibfield  {journal} {\bibinfo  {journal} {REVSTAT-Statistical Journal}\ }\textbf {\bibinfo {volume} {21}},\ \bibinfo {pages} {283} (\bibinfo {year} {2023})}\BibitemShut {NoStop}%
\bibitem [{\citenamefont {Hochreiter}\ and\ \citenamefont {Schmidhuber}(1997)}]{hochreiter1997long}%
  \BibitemOpen
  \bibfield  {author} {\bibinfo {author} {\bibfnamefont {S.}~\bibnamefont {Hochreiter}}\ and\ \bibinfo {author} {\bibfnamefont {J.}~\bibnamefont {Schmidhuber}},\ }\href@noop {} {\bibfield  {journal} {\bibinfo  {journal} {Neural Computation}\ }\textbf {\bibinfo {volume} {9}},\ \bibinfo {pages} {1735} (\bibinfo {year} {1997})}\BibitemShut {NoStop}%
\bibitem [{\citenamefont {Oreshkin}\ \emph {et~al.}(2020)\citenamefont {Oreshkin}, \citenamefont {Carpov}, \citenamefont {Chapados},\ and\ \citenamefont {Bengio}}]{oreshkin2019n}%
  \BibitemOpen
  \bibfield  {author} {\bibinfo {author} {\bibfnamefont {B.~N.}\ \bibnamefont {Oreshkin}}, \bibinfo {author} {\bibfnamefont {D.}~\bibnamefont {Carpov}}, \bibinfo {author} {\bibfnamefont {N.}~\bibnamefont {Chapados}}, \ and\ \bibinfo {author} {\bibfnamefont {Y.}~\bibnamefont {Bengio}},\ }in\ \href@noop {} {\emph {\bibinfo {booktitle} {8th International Conference on Learning Representations}}}\ (\bibinfo {year} {2020})\BibitemShut {NoStop}%
\bibitem [{\citenamefont {Challu}\ \emph {et~al.}(2023)\citenamefont {Challu}, \citenamefont {Olivares}, \citenamefont {Oreshkin}, \citenamefont {Ramirez}, \citenamefont {Canseco},\ and\ \citenamefont {Dubrawski}}]{challu2023nhits}%
  \BibitemOpen
  \bibfield  {author} {\bibinfo {author} {\bibfnamefont {C.}~\bibnamefont {Challu}}, \bibinfo {author} {\bibfnamefont {K.~G.}\ \bibnamefont {Olivares}}, \bibinfo {author} {\bibfnamefont {B.~N.}\ \bibnamefont {Oreshkin}}, \bibinfo {author} {\bibfnamefont {F.~G.}\ \bibnamefont {Ramirez}}, \bibinfo {author} {\bibfnamefont {M.~M.}\ \bibnamefont {Canseco}}, \ and\ \bibinfo {author} {\bibfnamefont {A.}~\bibnamefont {Dubrawski}},\ }in\ \href@noop {} {\emph {\bibinfo {booktitle} {Proceedings of the AAAI conference on artificial intelligence}}},\ \bibinfo {series and number} {\bibinfo {number} {6}}\ (\bibinfo {year} {2023})\ pp.\ \bibinfo {pages} {6989--6997}\BibitemShut {NoStop}%
\bibitem [{\citenamefont {Wu}\ \emph {et~al.}(2020)\citenamefont {Wu}, \citenamefont {Green}, \citenamefont {Ben},\ and\ \citenamefont {O'Banion}}]{wu2020deep}%
  \BibitemOpen
  \bibfield  {author} {\bibinfo {author} {\bibfnamefont {N.}~\bibnamefont {Wu}}, \bibinfo {author} {\bibfnamefont {B.}~\bibnamefont {Green}}, \bibinfo {author} {\bibfnamefont {X.}~\bibnamefont {Ben}}, \ and\ \bibinfo {author} {\bibfnamefont {S.}~\bibnamefont {O'Banion}},\ }\href@noop {} {\bibfield  {journal} {\bibinfo  {journal} {arXiv preprint arXiv:2001.08317}\ } (\bibinfo {year} {2020})}\BibitemShut {NoStop}%
\bibitem [{\citenamefont {Das}\ \emph {et~al.}(2023)\citenamefont {Das}, \citenamefont {Kong}, \citenamefont {Leach}, \citenamefont {Mathur}, \citenamefont {Sen},\ and\ \citenamefont {Yu}}]{das2023long}%
  \BibitemOpen
  \bibfield  {author} {\bibinfo {author} {\bibfnamefont {A.}~\bibnamefont {Das}}, \bibinfo {author} {\bibfnamefont {W.}~\bibnamefont {Kong}}, \bibinfo {author} {\bibfnamefont {A.}~\bibnamefont {Leach}}, \bibinfo {author} {\bibfnamefont {S.}~\bibnamefont {Mathur}}, \bibinfo {author} {\bibfnamefont {R.}~\bibnamefont {Sen}}, \ and\ \bibinfo {author} {\bibfnamefont {R.}~\bibnamefont {Yu}},\ }\href@noop {} {\bibfield  {journal} {\bibinfo  {journal} {arXiv preprint arXiv:2304.08424}\ } (\bibinfo {year} {2023})}\BibitemShut {NoStop}%
\bibitem [{\citenamefont {Hyndman}\ and\ \citenamefont {Athanasopoulos}(2018)}]{hyndman2018forecasting}%
  \BibitemOpen
  \bibfield  {author} {\bibinfo {author} {\bibfnamefont {R.~J.}\ \bibnamefont {Hyndman}}\ and\ \bibinfo {author} {\bibfnamefont {G.}~\bibnamefont {Athanasopoulos}},\ }\href@noop {} {\emph {\bibinfo {title} {Forecasting: principles and practice}}}\ (\bibinfo  {publisher} {OTexts},\ \bibinfo {year} {2018})\BibitemShut {NoStop}%
\bibitem [{\citenamefont {Panja}\ \emph {et~al.}(2024)\citenamefont {Panja}, \citenamefont {Chakraborty}, \citenamefont {Biswas},\ and\ \citenamefont {Deb}}]{panja2024stgcn}%
  \BibitemOpen
  \bibfield  {author} {\bibinfo {author} {\bibfnamefont {M.}~\bibnamefont {Panja}}, \bibinfo {author} {\bibfnamefont {T.}~\bibnamefont {Chakraborty}}, \bibinfo {author} {\bibfnamefont {A.}~\bibnamefont {Biswas}}, \ and\ \bibinfo {author} {\bibfnamefont {S.}~\bibnamefont {Deb}},\ }\href@noop {} {\bibfield  {journal} {\bibinfo  {journal} {arXiv preprint arXiv:2411.12258}\ } (\bibinfo {year} {2024})}\BibitemShut {NoStop}%
\bibitem [{\citenamefont {Koning}\ \emph {et~al.}(2005)\citenamefont {Koning}, \citenamefont {Franses}, \citenamefont {Hibon},\ and\ \citenamefont {Stekler}}]{koning2005m3}%
  \BibitemOpen
  \bibfield  {author} {\bibinfo {author} {\bibfnamefont {A.~J.}\ \bibnamefont {Koning}}, \bibinfo {author} {\bibfnamefont {P.~H.}\ \bibnamefont {Franses}}, \bibinfo {author} {\bibfnamefont {M.}~\bibnamefont {Hibon}}, \ and\ \bibinfo {author} {\bibfnamefont {H.~O.}\ \bibnamefont {Stekler}},\ }\href@noop {} {\bibfield  {journal} {\bibinfo  {journal} {International Journal of Forecasting}\ }\textbf {\bibinfo {volume} {21}},\ \bibinfo {pages} {397} (\bibinfo {year} {2005})}\BibitemShut {NoStop}%
\bibitem [{\citenamefont {Vovk}\ \emph {et~al.}(2005)\citenamefont {Vovk}, \citenamefont {Gammerman},\ and\ \citenamefont {Shafer}}]{vovk2005algorithmic}%
  \BibitemOpen
  \bibfield  {author} {\bibinfo {author} {\bibfnamefont {V.}~\bibnamefont {Vovk}}, \bibinfo {author} {\bibfnamefont {A.}~\bibnamefont {Gammerman}}, \ and\ \bibinfo {author} {\bibfnamefont {G.}~\bibnamefont {Shafer}},\ }\href@noop {} {\emph {\bibinfo {title} {Algorithmic learning in a random world}}},\ Vol.~\bibinfo {volume} {29}\ (\bibinfo  {publisher} {Springer},\ \bibinfo {year} {2005})\BibitemShut {NoStop}%
\bibitem [{Arctic climaric data source()}]{datasource}%
  \BibitemOpen
  Arctic climaric data source,\ \href@noop {} {}\bibinfo {howpublished} {\url{https://seklima.met.no/}},\ \bibinfo {note} {accessed: 2025-10-20}\BibitemShut {NoStop}%
\bibitem [{Code source for ``Forecasting precipitation in the Arctic using probabilistic machine learning informed by causal climate drivers''()}]{codesource}%
  \BibitemOpen
  Code source for ``Forecasting precipitation in the Arctic using probabilistic machine learning informed by causal climate drivers'',\ \href@noop {} {}\bibinfo {howpublished} {\url{https://zenodo.org/records/16728472}},\ \bibinfo {note} {accessed: 2025-10-20}\BibitemShut {NoStop}%
\bibitem [{\citenamefont {Kocarev}\ and\ \citenamefont {Parlitz}(1995)}]{kocarev1995general}%
  \BibitemOpen
  \bibfield  {author} {\bibinfo {author} {\bibfnamefont {L.}~\bibnamefont {Kocarev}}\ and\ \bibinfo {author} {\bibfnamefont {U.}~\bibnamefont {Parlitz}},\ }\href@noop {} {\bibfield  {journal} {\bibinfo  {journal} {Physical Review Letters}\ }\textbf {\bibinfo {volume} {74}},\ \bibinfo {pages} {5028} (\bibinfo {year} {1995})}\BibitemShut {NoStop}%
\bibitem [{\citenamefont {Kocarev}(2005)}]{kocarev2005complex}%
  \BibitemOpen
  \bibfield  {author} {\bibinfo {author} {\bibfnamefont {L.}~\bibnamefont {Kocarev}},\ }\href@noop {} {\emph {\bibinfo {title} {Complex dynamics in communication networks}}}\ (\bibinfo  {publisher} {Springer Science \& Business Media},\ \bibinfo {year} {2005})\BibitemShut {NoStop}%
\bibitem [{\citenamefont {Kocarev}(2013)}]{kocarev2013consensus}%
  \BibitemOpen
  \bibfield  {author} {\bibinfo {author} {\bibfnamefont {L.}~\bibnamefont {Kocarev}},\ }\href@noop {} {\emph {\bibinfo {title} {Consensus and synchronization in complex networks}}}\ (\bibinfo  {publisher} {Springer},\ \bibinfo {year} {2013})\BibitemShut {NoStop}%
\bibitem [{\citenamefont {Tang}\ \emph {et~al.}(2020)\citenamefont {Tang}, \citenamefont {Kurths}, \citenamefont {Lin}, \citenamefont {Ott},\ and\ \citenamefont {Kocarev}}]{tang2020introduction}%
  \BibitemOpen
  \bibfield  {author} {\bibinfo {author} {\bibfnamefont {Y.}~\bibnamefont {Tang}}, \bibinfo {author} {\bibfnamefont {J.}~\bibnamefont {Kurths}}, \bibinfo {author} {\bibfnamefont {W.}~\bibnamefont {Lin}}, \bibinfo {author} {\bibfnamefont {E.}~\bibnamefont {Ott}}, \ and\ \bibinfo {author} {\bibfnamefont {L.}~\bibnamefont {Kocarev}},\ }\href@noop {} {\bibfield  {journal} {\bibinfo  {journal} {Chaos: An Interdisciplinary Journal of Nonlinear Science}\ }\textbf {\bibinfo {volume} {30}},\ \bibinfo {pages} {063151} (\bibinfo {year} {2020})}\BibitemShut {NoStop}%
\end{thebibliography}%
\bibliographystyle{apsrev4-1}

\end{document}